\documentclass[11pt,a4paper]{article}
\pdfoutput=1

\usepackage{jcappub}
\usepackage{slashed}
\usepackage{braket}
\usepackage{hyperref}
\usepackage{float}
\usepackage{bbold}
\usepackage{amsmath}
\usepackage{amssymb}
\usepackage{graphicx}
\usepackage{xcolor}
\usepackage{mathrsfs}
\usepackage{comment}

\newcommand{\gadgetcode}{{\tt Gadget-2}}
\newcommand{\classcode}{{\tt class}}
\newcommand{\cambcode}{{\tt camb}}
\newcommand{\ngenic}{{\tt N-GenIC}}
\newcommand{\pkdgrav}{{\tt PKDGRAV3}}

\title{One line to run them all: SuperEasy massive neutrino linear response in $N$-body simulations}

\author[a]{Joe Zhiyu Chen,}
\author[a]{Amol Upadhye,} 
\author[a]{Yvonne Y. Y. Wong}

\affiliation[a]{Sydney Consortium for Particle Physics and Cosmology, School of Physics, The University of New South Wales, 
Sydney NSW 2052, Australia} 

\emailAdd{zhiyu.chen@unsw.edu.au} 
\emailAdd{a.upadhye@unsw.edu.au}
\emailAdd{yvonne.y.wong@unsw.edu.au}

\abstract{We present in this work a novel and yet extremely simple method for incorporating the effects of massive neutrinos in cosmological $N$-body simulations. This so-called ``SuperEasy linear response'' approach is based upon analytical solutions to the collisionless Boltzmann equation in the clustering and free-streaming limits, which are then connected by a rational function interpolation function with cosmology-dependent coefficients given by simple algebraic expressions of the cosmological model parameters.  The outcome is a {\it one-line modification} to the gravitational potential that requires  only the cold matter density contrast as a real-time input, and that can be incorporated into any $N$-body code with a Particle--Mesh component with no additional implementation cost.  To demonstrate its power, we implement the SuperEasy method in the publicly available \gadgetcode{} code, and show that for neutrino mass sums not exceeding $\sum m_\nu \simeq 1$~eV, the total matter and cold matter power spectra are in
sub-1\% and sub-0.1\% agreement with those from state-of-the-art linear response simulations in literature. Aside from its minimal implementation cost, compared with existing massive neutrino simulation methods, the SuperEasy approach has better memory efficiency, incurs no runtime overhead relative to a standard $\Lambda$CDM simulation, and requires no post-processing. The minimal nature of the method allows limited computational resources to be diverted to modelling other physical effects of interest, e.g., baryonic physics via hydrodynamics.}

\begin{document}

\begin{flushright}
	{\large \tt CPPC-2020-07}
\end{flushright}

\maketitle

\section{Introduction}

The absolute neutrino mass scale and mass hierarchy rank amongst the most actively researched topics today~\cite{Tanabashi:2018oca}. Flavour oscillation experiments using a variety of natural and man-made sources have pinned down the squared mass differences between generations~\cite{Abe:2008aa, Patrignani:2016xqp, deSalas:2017kay}.  Endpoint spectrum measurements~\cite{Kraus:2004zw, Osipowicz:2001sq, Esfahani:2017dmu} meanwhile are progressively pushing the maximum allowed effective electron neutrino mass  to sub-eV levels. Precision cosmology in the last decade  has also proven itself to be an invaluable laboratory for neutrino physics~\cite{Lesgourgues:2006nd, Wong:2011ip}: Current observations of the  cosmic microwave background (CMB) anisotropies and the large-scale structure already constrain the neutrino sum to $\Sigma m_{\nu} \lesssim {\cal O}(0.5)\, \text{eV}$, depending on model assumptions~\cite{Aghanim:2018eyx,Upadhye:2017hdl}; the next generation of surveys~\cite{Abell:2009aa, Laureijs:2011gra, Abazajian:2016yjj} has the potential to improve this limit by tenfold~\cite{Chudaykin:2019ock}. Enabling this advancement  therefore forms the wider goal of this work.

In the cosmological context, nonzero neutrino masses have two main effects on the large-scale structure distribution. 
Firstly, while sub-eV mass neutrinos are nonrelativistic today and count towards the present-day mean total matter density $\bar{\rho}_{\rm m,0}$, 
at early times they are ultrarelativistic and behave like radiation.  Thus, relative to the massless case, a massive neutrino cosmology will, for the same $\bar{\rho}_{\rm m,0}$,  generally suffer a delayed  transition from radiation to matter domination.  This in turn delays the onset of perturbation growth, leading ultimately to a suppression of power in the present-day large-scale matter power spectrum  on those (small) scales that entered the horizon during radiation domination.

The second effect of neutrino masses on structure formation stems from the neutrinos' free-streaming behaviour. Even after they have become nonrelativistic at late times, a large thermal velocity that scales inversely with the particle mass enables cosmological neutrinos to avoid being gravitationally captured on small length scales in an efficient manner.  This inefficiency adds to the small-scale power suppression in the large-scale matter power spectrum discussed above, and is commonly understood to be the main cause of late-time scale-dependent growth in massive neutrino cosmologies.
 
From a theoretical and computational perspective, the effects of neutrino masses on cosmology are well understood within the framework of relativistic linear perturbation theory~\cite{Lesgourgues:2012uu}, and publicly available linear cosmological Boltzmann codes such as \cambcode{}~\cite{Lewis:1999bs} or  \classcode{}~\cite{Blas:2011rf} are able to predict their small-scale signature in the linear matter power spectrum to sub-precent precision in a matter of seconds.  However, as the matter density contrasts grow to exceed unity, nonlinear dynamics kick in progressively from small to large scales.
The challenge for us, therefore, is to quantify these same small-scale effects  under {\it nonlinear} evolution, through which to compute the nonlinear matter power spectrum (and other observables) in massive neutrino cosmologies to the same sub-percent accuracy as its linear counterpart and in as efficient a manner as possible.

The study of nonlinear cosmic structure formation traditionally belongs in the realm of numerical $N$-body simulations. Such simulations are however computationally expensive, a problem that  is exacerbated by the addition of massive neutrinos, because of the difficulty in modelling their large thermal velocity. On the one hand,  if treated as a fluid on a grid~\cite{Dakin:2017idt}, the Courant condition limits the fluid elements to not traverse more than a single grid cell per time step; at high redshifts, this slows down the simulation considerably as highly relativistic neutrinos require extremely fine time stepping. 
On the other hand, modelling neutrinos as a collection of simulation particles~\cite{Brandbyge:2008rv, Viel:2010bn, Villaescusa-Navarro:2013pva, Adamek:2017uiq, Bayer:2020tko} presents a different set of numerical challenges:  to adequately sample the neutrino phase space distribution requires at minimum as many particles as the cold matter species, lest a large velocity dispersion coupled with mild clustering eventuate in a poor signal-to-noise ratio~\cite{Banerjee:2018bxy, Elbers:2020lbn}.

To circumvent these issues, one efficient approach has emerged in the past decade  that combines a perturbative  neutrino component  with a full $N$-body realisation for the cold matter~\cite{Brandbyge:2008js}.  The rationale for this type of split modelling follows from the observation that fast-moving neutrinos cannot cluster strongly, so that their spatial density contrast should remain in a regime amenable  to perturbative treatments.  Existing implementations of such a split approach generally employ an Eulerian perturbation theory to track neutrino density perturbations on the mesh of a Particle--Mesh (PM) gravity solver, thereby allowing neutrinos to contribute to the gravitational potentials
despite not being represented as particles in the simulation volume~\cite{Brandbyge:2008js, Partmann:2020qzb, AliHaimoud:2012vj, Bird:2018all,Tram:2018znz}. 

Explorations of this class of split simulations have so far been limited to linearised neutrino equations of motion, and  we further distinguish two variants:
\begin{enumerate}
\setlength\itemsep{-0.1em} 
\item Linear neutrino perturbations~\cite{Brandbyge:2008js,Tram:2018znz}, obtained from solving a {\it completely} linearised system, including linearised equations of motion for the cold matter;  in practice this can be achieved either by running \cambcode{} or \classcode{} concurrently with the simulation, or by pre-computing the linear neutrino perturbations at the required simulation time steps; 
and 
\item Linear response~\cite{AliHaimoud:2012vj, Bird:2018all}, where the linearised neutrino equations may respond to an external potential sourced by {\it nonlinear} cold matter clustering as determined by the $N$-body component of the simulation. 
\end{enumerate}
For small enough neutrino masses ($m_\nu \lesssim 0.5$ eV), both variants are able to reproduce neutrino free-streaming effects on the cold matter power spectrum to sub-percent agreement with a full $N$-body simulation~\cite{AliHaimoud:2012vj}.  Neither has been designed to yield the correct neutrino densities --- certainly not the linear neutrino perturbation variant --- although analyses of spherical systems~\cite{Ringwald:2004np} suggest that linear response estimates may come within a factor of a few of the true expectations even on cluster and galactic scales.%
\footnote{There exists also a class of simulations that relies on post-processing to remap a cold matter-only simulation to describe a massive neutrino cosmology~\cite{Lawrence:2017ost,Partmann:2020qzb}.  The post-processing typically involves rescaling some aspects of the original simulation by quantities determined from linear perturbation theory (e.g.,  linear growth factor ratios of the original and the target cosmologies).  As with simulations that use linear neutrino perturbations~\cite{Brandbyge:2008js,Tram:2018znz}, this class of simulations is inherently unable to predict the neutrino perturbations under nonlinear dynamics.}

The present work is part of a set of two papers in which we explore further the linear response approach.  In paper 1 (this work), we take as a starting point the same perturbation theory as~\cite{AliHaimoud:2012vj}, and proceed to streamline it into the titular ``SuperEasy'' method.  In the companion paper 2~\cite{Chen:2020}, we explore a ``semi-Lagrangian'', multi-fluid implementation of linear response that dovetails with the so-called hybrid schemes, wherein neutrinos are converted from Eulerian perturbations sitting on the mesh points to a particle representation once their velocities drop to suitably low and tractable values~\cite{Brandbyge:2009ce, Bird:2018all}. 

Specifically, the SuperEasy linear response method builds upon closed-form solutions for the neutrino density contrast's response to a given cold matter perturbation in the clustering and free-streaming limits~\cite{Ringwald:2004np}.  With these limiting solutions in hand, a simple rational function-interpolated response function in terms of the neutrino free-streaming scale emerges for the total matter density contrast across the scales of interest, allowing {\it any} cold matter spatial density contrast to be instantaneously mapped to its total matter counterpart.   Then, accounting for the effects of neutrino masses in a PM $N$-body simulation becomes as simple as a one-line modification to the Poisson equation on the PM-grid to incorporate the said interpolated response function.

  In terms of computational costs, the SuperEasy linear response method is as cheap as it gets.  It incurs no runtime overhead compared with a standard $\Lambda$CDM simulation, and, relative to other methods of comparative runtime performance~(e.g., \cite{Partmann:2020qzb}),  the SuperEasy method requires no post-processing, and has the added benefit of being able to predict the nonlinear neutrino density contrast to some degree of accuracy. The SuperEasy method  is also easily generalisable to handle multiple non-degenerate neutrino masses, again at minimal implementation and practically no additional computational costs.  
  The slim footprint of the method allows limited computational resources to be diverted to achieving higher resolutions in the cold matter sector, or, when the inclusion of baryonic physics necessitates it, to  the implementation of hydrodynamics.
  
The paper is organised as follows.  We review  in section~\ref{Sec:BoltzmannEqn} the collisionless Boltzmann equation for nonrelativistic massive neutrinos, and formally solve it in the linear, Newtonian limit to arrive at the integral linear response function used in the simulations of~\cite{AliHaimoud:2012vj,Bird:2018all}. In section~\ref{Sec:Superfast}, we further reduce this integral response function in the free-streaming and clustering limits into closed forms, and demonstrate that a simple interpolation function, which forms the basis of the SuperEasy linear response method, 
is able to capture the linear total matter power spectrum output of \classcode{} to sub-percent accuracy.    The $N$-body implementation of the SuperEasy method  is presented in section~\ref{Sec:Nbody}, along with  results of our convergence tests in section~\ref{Sec:Results}. We contrast the predictions of the SuperEasy method with those of other $N$-body approaches in section~\ref{sec:comparison}, and assess the general validity of linear response methods in section~\ref{subsec:neutrino_density_contrast_from_linear_response}.
 Section~\ref{Sec:Conclusions} contains our conclusions.   A generalisation of the SuperEasy formalism to multiple non-degenerate neutrino species can be found in appendix~\ref{Sec:ExtHierarchy}.


\section{Collisionless Boltzmann equation and linear response}
\label{Sec:BoltzmannEqn}

Consider the relic neutrinos as a nonrelativistic gas of collisionless gravitating particles in an expanding background.  On deeply subhorizon scales, it is convenient to use the comoving coordinates~$\vec{x}$, the conformal time~$\tau$, and a comoving momentum defined as $\vec{p} \equiv a m_\nu \dot{\vec{x}}$,  where $\cdot \equiv \partial/\partial \tau$, $a$ is the scale factor, and $m_\nu$ is the neutrino mass.  Then, the  phase space density of the neutrino gas \(f(\vec{x},\vec{p},\tau)\) can be defined via the number of particles in an infinitesimal phase space volume, 
$\mathrm{d}N \equiv f(\vec{x},\vec{p},\tau) \mathrm{d}^3x \, \mathrm{d}^3p$,
and the collisionless Boltzmann or Vlasov equation, 
\begin{equation}
	\frac{\mathrm{d} f}{\mathrm{d}\tau} \equiv \frac{\partial f}{\partial \tau} + \frac{\vec{p}}{am_\nu} \cdot \nabla_{\!\vec{x}} \, f - am_\nu \nabla_{\!\vec{x}} \, \phi \cdot \nabla_{\!\vec{p}} \, f = 0  
	\label{Eq:BoltzmannEq}
\end{equation}
governs its time evolution in the presence of a Newtonian gravitational potential  $\phi = \phi(\vec{x},\tau)$.

The gravitational potential~$\phi$ is itself determined by  the spatial  fluctuations of matter density via the Poisson equation,
 \begin{equation}
 \begin{aligned}
 \nabla^2_{\vec x} \, \phi(\vec{x},\tau) =& \, 4 \pi G\, a^2(\tau) \, \bar{\rho}_{\rm m} (\tau)\delta_{\rm m} (\vec{x},\tau) \\
 = & \, \frac{3}{2} {\cal H}^2(\tau) \, \Omega_{\rm m}(\tau)  \, \delta_{\rm m} (\vec{x},\tau).
 \label{eq:poisson}
 \end{aligned}
 \end{equation}
Here, $\bar{\rho}_{\rm m} = \bar{\rho}_{\rm cb}+ \bar{\rho}_\nu$ is the mean total matter density which we take to consist of a combined cold matter (cold dark matter and baryons; 	``cb'') and a neutrino (``$\nu$'' ) component, ${\cal H} \equiv(1/a) ({\rm d}a /{\rm d} \tau)$ is the conformal Hubble expansion rate, 
$\Omega_{\rm m}(\tau) \equiv \bar{\rho}_{\rm m}(\tau)/\rho_{\rm crit}(\tau)$  the time-dependent reduced matter density, 
 and  
 \begin{equation}
 \delta_{\rm m}= f_{\rm cb} \delta_{\rm cb} + f_\nu \delta_{\nu}
 \label{eq:totalcontrast}
 \end{equation}
  is the total matter density contrast, in which the individual cb and $\nu$ density contrasts are weighted by $f_{\rm cb}  \equiv \bar{\rho}_{\rm cb}/\bar{\rho}_{\rm m}$ and $f_{\nu}  \equiv \bar{\rho}_{\nu}/\bar{\rho}_{\rm m}$ respectively.
The neutrino density contrast~$\delta_{\nu}$ can  be constructed from the phase space density \(f(\vec{x},\vec{p},\tau)\) via 
\begin{equation}
\delta_{\nu}  (\vec{x},\tau) \equiv \frac{\rho_{\nu}(\vec{x},\tau)-\bar{\rho}_\nu(\tau)}{\bar{\rho}_\nu(\tau)} = \frac{\int {\rm d}^3 p\, f(\vec{x},\vec{p},\tau)-\bar{f}(p)}{\int {\rm d}^3 p \, \bar{f}(p)},
\label{eq:nudensitycontrast}
\end{equation} 
where $\bar{f}(p)$ is the phase space density of the homogeneous background given, for neutrinos, by the relativistic Fermi--Dirac distribution,  $\bar{f}(p)= [1 + \exp(p/T_{\nu,0})]^{-1}$, with present-day temperature $T_{\nu,0}$.  For the purpose of laying down the linear response framework in sections~\ref{Sec:BoltzmannEqn} and~\ref{Sec:Superfast}, the cold matter density contrast~$\delta_{\rm cb} \equiv \rho_{\rm cb} (\vec{x},\tau)/ \bar{\rho}_{\rm cb} (\tau)-1$ can be  treated as a given external function.

 
 \subsection{Integral linear response}
 
To obtain a formal solution to this Poisson--Boltzmann system~\eqref{Eq:BoltzmannEq}--\eqref{eq:poisson}, we  follow the prescription of~\cite{Bertschinger:1993xt} and Fourier-transform
equation~\eqref{Eq:BoltzmannEq} according to $A(\vec{k}) = {\cal F}[A(\vec{x})] \equiv \int_{-\infty}^{\infty} A(\vec{x}) \, e^{-i \vec{k} \cdot \vec{x}} \, \mathrm{d}^3x$.  Rewriting the transformed equation in terms of a superconformal time variable defined as  \(s \equiv \int a^{-1} \mathrm{d}\tau\), we find
\begin{equation}
	\frac{\partial {f}}{\partial s} + \frac{i \, \vec{k} \cdot \vec{p}}{m_\nu} {f}-  i m_\nu a^2 (\vec{k} \, {\phi} \ast \nabla_{\!\vec{p}} \, {f}) = 0 \, ,
	\label{Eq:FourierBoltzmann}
\end{equation}
where $A(\vec{k})* B(\vec{k}) \equiv \int {\rm d}^3 k' \, A(\vec{k}') B(\vec{k} - \vec{k}')$ denotes a convolution product.  Observe that the convolution term is also the only instance of $\vec{k}$-mode coupling in equation~\eqref{Eq:FourierBoltzmann}.  Therefore, we simplify it first by linearisation, i.e.,
\begin{equation}
	\begin{aligned}
		\vec{k} \, {\phi} \ast \nabla_{\!\vec{p}} \, {f} \simeq & \, \vec{k} \, {\phi} \ast \nabla_{\!\vec{p}} \left(\bar{f}(p) \, \delta_D^{(3)}(\vec{k})\right) \\
		= & \, \vec{k} \, {\phi} \cdot \nabla_{\!\vec{p}} \, \bar{f}(p) \, ,
		\label{eq:linearisation}
	\end{aligned}
\end{equation}
where $\delta^{(3)}_D(\vec{k})$ denotes a 3-dimensional Dirac delta distribution in $k$-space,   Physically, the linearisation scheme~\eqref{eq:linearisation} corresponds to assuming that gravitational clustering does not significantly distort the neutrino phase space density from its homogeneous expectation, 
\begin{equation}
 |\nabla_{\!\vec{p}} \, (f - \bar{f})| \ll |\nabla_{\!\vec{p}} \, \bar{f}|,
 \label{eq:linearisationcondition}
 \end{equation}
  an assumption that should hold as long as the neutrino density contrast remains well below unity.  We shall revisit this issue of linearisation in section~\ref{subsec:neutrino_density_contrast_from_linear_response}.

Upon linearisation, the Fourier-space Boltzmann equation~\eqref{Eq:FourierBoltzmann} now reduces to a first-order inhomogeneous differential equation of the form
\begin{equation}
	\frac{\partial {f}}{\partial s} + \frac{i \vec{k} \cdot \vec{p}}{m_\nu} {f} - i m_\nu a^2 (\vec{k} \, {\phi} \cdot \nabla_{\!\vec{p}} \, \bar{f}) = 0 \, .
	\label{Eq:LinearFourierBoltzmann}
\end{equation}
Treating ${\phi}(\vec{k},s)$ as an external potential, equation~\eqref{Eq:LinearFourierBoltzmann} is formally solved by~\cite{Bertschinger:1993xt} 
\begin{equation}
	\begin{aligned}
		{f}(\vec{k},\vec{p},s) =& \, {f}(\vec{k},\vec{p},s_{\rm i}) \, \text{exp} \! \left(-\frac{i\vec{k}\cdot\vec{p}}{m_{\nu}} (s-s_{\rm i})\right) \\
		& \hspace{1.5cm} + i m_{\nu} \vec{k} \cdot \nabla_{\!\vec{p}} \, \bar{f} \int_{s_{\rm i}}^s \mathrm{d}s' \, a^2(s') \, {\phi}(\vec{k},s') \, \text{exp} \! \left(-\frac{i\vec{k}\cdot\vec{p}}{m_{\nu}} (s-s') \right) \, .
	\end{aligned}
	\label{Eq:GenSolution}
\end{equation}
Physically, the solution to the homogeneous equation (i.e., the first term) corresponds to redistribution of the initial conditions at $s= s_{\rm i}$ in the neutrino sector via free-streaming.  In contrast, the inhomogeneous solution (the second term) represents the neutrinos' response to the formally external gravitational potential --- hence the name ``linear response''.

Then, integrating~\eqref{Eq:GenSolution} in momentum as per equation~\eqref{eq:nudensitycontrast},  we  find the neutrino density contrast ${\delta}_\nu(\vec{k},s)$ at the mode~$\vec{k}$ and time~$s$ to be~\cite{Bertschinger:1993xt} 
\begin{equation}
	\begin{aligned}
		{\delta}_{\nu}(\vec{k},s) \simeq - k^2 \int_{s_{\rm i}}^s \mathrm{d}s' \, a^2(s') \, \phi(\vec{k},s') \, (s-s') \, F\left[\frac{T_{\nu,0}k(s-s')}{m_{\nu}}\right] \,  ,
	\end{aligned}
	\label{Eq:MasterNeutrinoEquation}
\end{equation}
where 
\begin{equation}
	\begin{aligned}
		F(q) \equiv \frac{m_\nu}{\bar{\rho}_{\nu}(s)} \int \mathrm{d}^3p \,  \bar{f}(p) \, \text{exp}\left(- i \vec{q} \cdot \vec{p}/T_{\nu,0} \right)  \, 
	\end{aligned}
	\label{Eq:FDef}
\end{equation}
is effectively a Fourier transform of the homogeneous phase density~$\bar{f}(p)$ in the comoving momentum variable~$\vec{p}$ to a new variable~$\vec{q}$.   For a relativistic Fermi--Dirac distribution, $F(q)$ evaluates to
\begin{equation}
\begin{aligned}
F(q) & = \frac{4}{3\zeta(3)} \sum_{n=1}^\infty (-1)^{n+1} \frac{n}{(n^2 + q^2)^2}  \\
& = \frac{i}{12\zeta(3)q }  \Bigg[ \psi^{(1)}\left(\frac{1+iq}{2}\right) - \psi^{(1)} \left(\frac{1-iq}{2}\right) + \psi^{(1)}\left(-\frac{iq}{2}\right) - \psi^{(1)}\left(\frac{iq}{2}\right) \Bigg] ,
\label{eq:series}
\end{aligned}
\end{equation}
with the  Riemann zeta function  \(\zeta(3) \simeq 1.202\), and  $\psi^{(i)}$ is a polygamma function of order~$i$.
 Observe that in writing equation~\eqref{Eq:MasterNeutrinoEquation} we have also assumed the free-streaming redistribution term in equation~\eqref{Eq:GenSolution} to integrate approximately to the homogeneous background density $\bar{\rho}_\nu(s)$, such that the final ${\delta}_{\nu}(\vec{k},s)$ depends only on its response to a formally external gravitational source.

Equation~\eqref{Eq:MasterNeutrinoEquation} is the starting point of all existing linear response calculations of nonrelativistic gravitational neutrino clustering in the literature.  To distinguish it from the SuperEasy linear response method proposed in this work, we shall refer to it as ``integral linear response''  because of the remaining integral over time.
Integral linear response~\eqref{Eq:MasterNeutrinoEquation} has been previously applied to the investigation of neutrino clustering around cosmic string loops~\cite{Brandenberger:1987kf}, dark matter halos~\cite{Abazajian:2004zh}, and more recently in $N$-body simulations of large-scale structure~\cite{AliHaimoud:2012vj}.   We defer to section~\ref{Sec:ComparisonILR} our discussion of the $N$-body implementation of integral linear response as laid down in~\cite{AliHaimoud:2012vj},  opting to demonstrate first  in the next section how equation~\eqref{Eq:MasterNeutrinoEquation}  can be further manipulated to form the basis of our SuperEasy method.


\section{SuperEasy linear response}
\label{Sec:Superfast}

The basis of our SuperEasy linear response method lies in the realisation that the remaining time-integral in equation~\eqref{Eq:MasterNeutrinoEquation} is in fact analytically soluble in the large $k$ (``free-streaming'') limit~\cite{Ringwald:2004np}. In the opposite small $k$ (``clustering'') limit, there are likewise strong physical grounds upon which to predict the evolution of ${\delta}_{\nu}$.  Here, we discuss first the solutions of integral~\eqref{Eq:MasterNeutrinoEquation} in the large and small~$k$ limits, before introducing in section~\ref{sec:superfastpotential} the centrepiece of the SuperEasy linear response method based on these limiting behaviours.


\subsection{Free-streaming limit}

Beginning with equation~\eqref{Eq:MasterNeutrinoEquation}, 
we first split up the time-integral into ${\delta}_\nu = \int u\,  {\rm d}v$, with
\begin{equation}
	\begin{aligned}
		u \equiv & \, - k^2\, a^2(s') \, \phi (\vec{k}, s') \, ,\\
		\mathrm{d}v \equiv & \, (s-s') \, F\left[\frac{T_{\nu, 0}k(s-s')}{m_{\nu}}\right] \, \mathrm{d}s' \, .
	\end{aligned}
	\label{Eq:MasterEqByParts}
\end{equation}
Using $F(q)$ from equation~\eqref{eq:series} for $q \equiv T_{\nu,0} k(s-s')/m_\nu$, 
the latter evaluates to
\begin{equation}
\begin{aligned}
 v  = \frac{2}{3\zeta(3)} \left(\frac{m_{\nu}}{k T_{\nu,0}}\right)^2 \sum_{n=1}^\infty (-1)^{n+1} \frac{n}{n^2 + q^2} = \frac{2}{3\zeta(3)} \left(\frac{m_{\nu}}{k T_{\nu,0}}\right)^2  \mathscr{G}(q),
 \label{eq:v}
 \end{aligned}
\end{equation}
with 
\begin{equation}
\begin{aligned}
\mathscr{G}(q) \equiv - \frac{1}{4}\Bigg[ \psi^{(0)}\left(\frac{1+iq}{2}\right) + \psi^{(0)} \left(\frac{1-iq}{2}\right) + \psi^{(0)}\left(-\frac{iq}{2}\right) + \psi^{(0)}\left(\frac{iq}{2}\right) \Bigg].  
\label{eq:scriptf}
\end{aligned}
\end{equation}
It then follows from integration by part  (i.e., ${\delta}_\nu = u v - \int v \, {\rm d} u$) that
\begin{equation}
\begin{aligned}
{\delta}_\nu (\vec{k}, s) = &  \, - \frac{2}{3 \zeta(3)} \left(\frac{m_{\nu}}{T_{\nu,0}}\right)^2 \left\{ a^2(s') \, \phi(\vec{k},s')  \left. \mathscr{G} \left[ \frac{T_{\nu,0} k (s-s')}{m_\nu}\right]\right|_{s_{\rm i}}^s \right. \\
& \hspace{40mm}
-\left.  \int_{s_{\rm i}}^s {\rm d} s' \,  \frac{\mathrm{d}(a^2 \phi)}{\mathrm{d}s'} \, \mathscr{G} \left[ \frac{T_{\nu,0} k (s-s')}{m_\nu}\right] \right\} 
\end{aligned}
\label{Eq:ByPartsComponents}
\end{equation}
is formally equivalent to the solution~\eqref{Eq:MasterNeutrinoEquation}.

To simplify the expression~\eqref{Eq:ByPartsComponents},  observe first of all that the function $\mathscr{G}(q)$ is positive and finite: it approaches $\ln(2)$ as $q \to 0$, remains fairly flat up to $q  \simeq 1$, and drops off quickly to zero beyond $q \simeq 1$.   Then, 
because the initial time \(s_{\rm i}\) can be set to an arbitrarily distant past where \(a(s_i) \ll a(s)\) and \({\delta}_{\rm m}(s_i) \ll {\delta}_{\rm m}(s)\), we can immediately approximate the first term of equation~\eqref{Eq:ByPartsComponents} with
\begin{equation}
a^2(s') \, \phi (\vec{k},s')  \left. \mathscr{G} \left[ \frac{T_{\nu,0} k (s-s')}{m_\nu}\right]\right|_{s_{\rm i}}^s  \simeq \ln(2)\,  a^2(s) \, \phi(\vec{k},s). 
\label{eq:term1}
\end{equation}
Secondly,  the time-integral in equation~\eqref{Eq:ByPartsComponents} may be eliminated in the free-streaming limit by the following argument.  It is straightforward to establish that $\mathrm{d}(a^2 \phi)/\mathrm{d}s'$
appearing in the integrand is a monotonically increasing function of time.  Combined with the flatness of $\mathscr{G}(q)$  at $q \lesssim 1$,  we can conclude that the dominant contribution to the time-integral must come from the time interval immediately before the present time, $\Delta s \equiv s-s' \simeq m_\nu/(k T_{\nu,0})$, and on this basis approximate the time-integral as
\begin{equation}
\begin{aligned}
 \int_{s_{\rm i}}^s {\rm d} s' \,  \frac{\mathrm{d}(a^2 \phi)}{\mathrm{d}s'} \, \mathscr{G} \left[ \frac{T_{\nu,0} k (s-s')}{m_\nu}\right] & \simeq  \ln(2)  \left. a^2(s')\, \phi(s') \right|^s_{s-\Delta s}\\
 & \sim  \ln(2) \, \frac{{\rm d} (a^2 \phi)}{ {\rm d} s} \frac{m_\nu}{k T_{\nu,0}}.
 \label{eq:term2}
 \end{aligned}
\end{equation}
Comparing equations~\eqref{eq:term1} and~\eqref{eq:term2}, we see immediately that if the condition
\begin{equation}
\frac{1}{a^2 \phi}\frac{{\rm d} (a^2 \phi)}{ {\rm d} s} \frac{m_\nu}{k T_{\nu,0}} \ll 1
\label{eq:condition}
\end{equation}
is satisfied, then  the time-integral in equation~\eqref{Eq:ByPartsComponents} must make but a negligible contribution to $\delta_{\nu}(\vec{k},s)$, and can therefore be dropped in our calculations.

To further interpret the condition~\eqref{eq:condition}, we note that the relative rate of change of $a^2 \phi$
with respect to the superconformal time are typically of order $a {\cal H}$. 
Thus, the condition~\eqref{eq:condition} under which we may neglect the time-integral contribution to~${\delta}_\nu(\vec{k},s)$ 
is physically equivalent to taking the free-streaming limit,
\begin{equation}
{\cal H}  \ll  \frac{k T_{\nu}}{m_\nu} \, ,
\label{Eq:FreeStreamCondition}
\end{equation}
where $T_{\nu}(s) = T_{\nu,0}/a(s)$ is the temperature of the neutrino population at a given time~$s$, and $k T_{\nu}/m_\nu$ corresponds to the population's characteristic thermal velocity.

Then, returning to equation~\eqref{Eq:ByPartsComponents} and applying to it the approximations discussed above as well as the Poisson equation~\eqref{eq:poisson},
  we find in the free-streaming limit~\cite{Ringwald:2004np}
\begin{equation}
\begin{aligned}
{\delta}^{\rm FS}_\nu(\vec{k},s) \, \simeq \, &  \, \frac{3}{2}  {\cal H}^2(s)\,
\Omega_{\rm m}(s)\,
 \frac{2 \ln(2) }{3\zeta(3)} \left(\frac{m_{\nu}}{k T_{\nu}} \right)^2   {\delta}_{\rm m}(\vec{k},s).
\end{aligned}
\label{Eq:IntegralStepFreeStream}
\end{equation} 
Identifying a free-streaming wave number~\(k_{\text{FS}}\) and an associated sound speed~\(c_{\nu}\)~\cite{Ringwald:2004np}
\begin{eqnarray}
k_{\text{FS}} (s)&\equiv & \, \sqrt{\frac{(3/2)\, {\cal H}^2(s)\, \Omega_{\rm m}(s)}{c_{\nu}^2(s)}}  \simeq 1.5 \, \sqrt{a(s)\, \Omega_{\rm m,0}} \left( \frac{m_\nu}{\rm eV}\right)\, h/{\rm Mpc}\, , \label{Eq:kFs} \\
c_{\nu} (s) &\equiv & \, \frac{T_{\nu}(s)}{m_{\nu}} \sqrt{\frac{3 \, \zeta(3)}{2 \ln(2)}} \simeq \frac{81}{a(s)} \left(\frac{\rm eV}{m_\nu} \right)\, {\rm k m/s}
\, ,
\label{Eq:FreeStreamingScale} 
\end{eqnarray}
where $\Omega_{\rm m,0}$ is the present-day reduced matter density, we finally arrive at the free-streaming solution~\cite{Ringwald:2004np}
\begin{equation}
{\delta}^{\rm FS}_{\nu}(\vec{k},s)  \simeq \frac{k^2_{\text{FS}}(s)}{k^2} \, {\delta}_{\rm m}(\vec{k},s) \, ,
\label{Eq:FreeStreamDeltaNuSolution}
\end{equation}
and we can remap the free-streaming condition~\eqref{Eq:FreeStreamCondition} equivalently  to the requirement $k \gg k_{\rm FS}$.

For a total matter content consisting only of cold matter and neutrinos, equation~\eqref{Eq:FreeStreamDeltaNuSolution} may be rewritten  in terms of the cb density contrast as
\begin{equation}
 {\delta}^{\rm FS}_{\nu}(\vec{k},s)  \simeq \frac{k^2_{\text{FS}} (s)(1-f_{\nu})}{k^2 - k^2_{\text{FS}} (s) f_{\nu}} \, {\delta}_{\text{cb}}(\vec{k},s) \, 
\label{Eq:NeutrinoDeltaInCB}
\end{equation}
following equation~\eqref{eq:totalcontrast}.  Given $k \gg k_{\rm FS}$, the expression~\eqref{Eq:NeutrinoDeltaInCB} shows that the neutrino density contrast is always suppressed by $k^{-2}$ relative to its cold matter counterpart,  reflecting the neutrinos' increasing tendency to cluster less efficiently on smaller length scales.


\subsection{Clustering limit}
\label{sec:clustering}

To investigate the  behaviour of the integral linear response function~\eqref{Eq:MasterNeutrinoEquation} in the opposite,  \(k \ll k_{\text{FS}}\) limit, we formally set the argument of $F(q)$ to zero, so that $F(0) = 1$ by the definition~\eqref{Eq:FDef}, and 
\begin{equation}
	\begin{aligned}
	{\delta}^{\rm C}_{\nu}(\vec{k}, s)\simeq \, - k^2 \int_{s_{\rm i}}^s {\rm d} s'\, a^2(s') \, \phi(\vec{k}, s') \, (s-s') \,  
	\end{aligned}
	\label{Eq:DeltaNuClusteringForm}
\end{equation}
is the solution of the neutrino density contrast in the clustering (``C'') limit.

Without specifying the time-dependences of $a(s)$ and $\phi(s)$, equation~\eqref{Eq:DeltaNuClusteringForm} cannot be simplified any further.  However, we note that equation~\eqref{Eq:DeltaNuClusteringForm} is in fact a solution to the second-order differential equation 
\begin{equation}
		\frac{\partial ^2 {\delta}_{\nu}}{\partial s^2} = - k^2 \, a^2(s) \, \phi(\vec{k},s) \, ,
	\label{Eq:DeltaNuClusteringCDM}
\end{equation}
the same equation of motion that determines the cold matter density contrast ${\delta}_{\text{cb}}$ at linear order~\cite{Bernardeau:2001qr}, up to initial conditions.  Since we generally expect the clustering limit to reside in the linear regime, together with the assumption of adiabatic initial conditions we can reasonably conclude from the above observation that
 \begin{equation}
{\delta}^{\rm C}_{\nu}(\vec{k},s)\simeq  {\delta}_{\text{cb}}(\vec{k},s)  \simeq  {\delta}_{\text{m}}(\vec{k},s) 
	\, ,
	\label{Eq:ClusteringDeltaNuSolution}
\end{equation}
 a result that can be readily verified by the transfer function outputs of linear cosmological Boltzmann codes such as \cambcode{}~\cite{Lewis:1999bs} or \classcode{}~\cite{Blas:2011rf}.


\subsection{Interpolating between limits}

\begin{figure}
	\begin{center}
		\includegraphics[width=75mm]{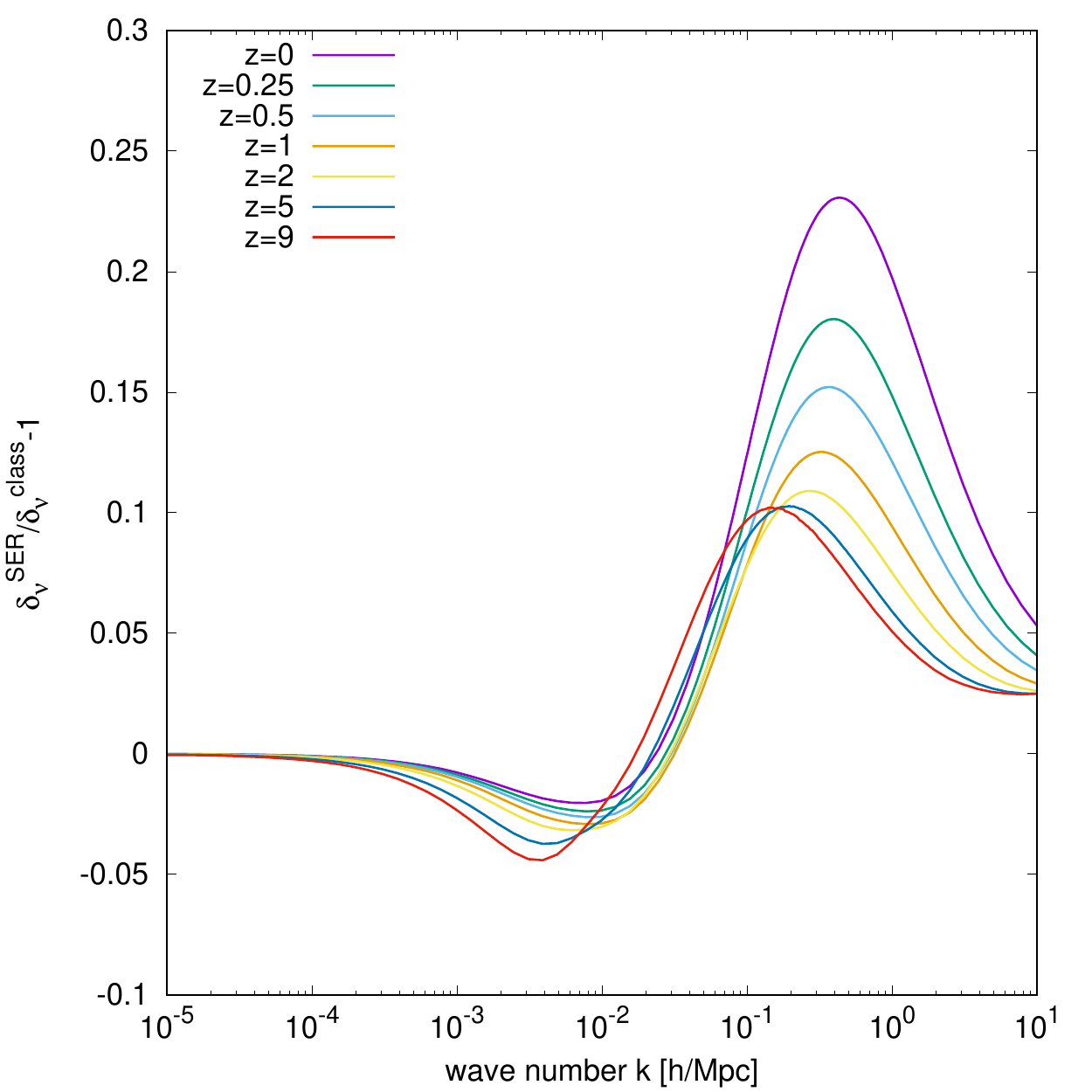}
		\includegraphics[width=75mm]{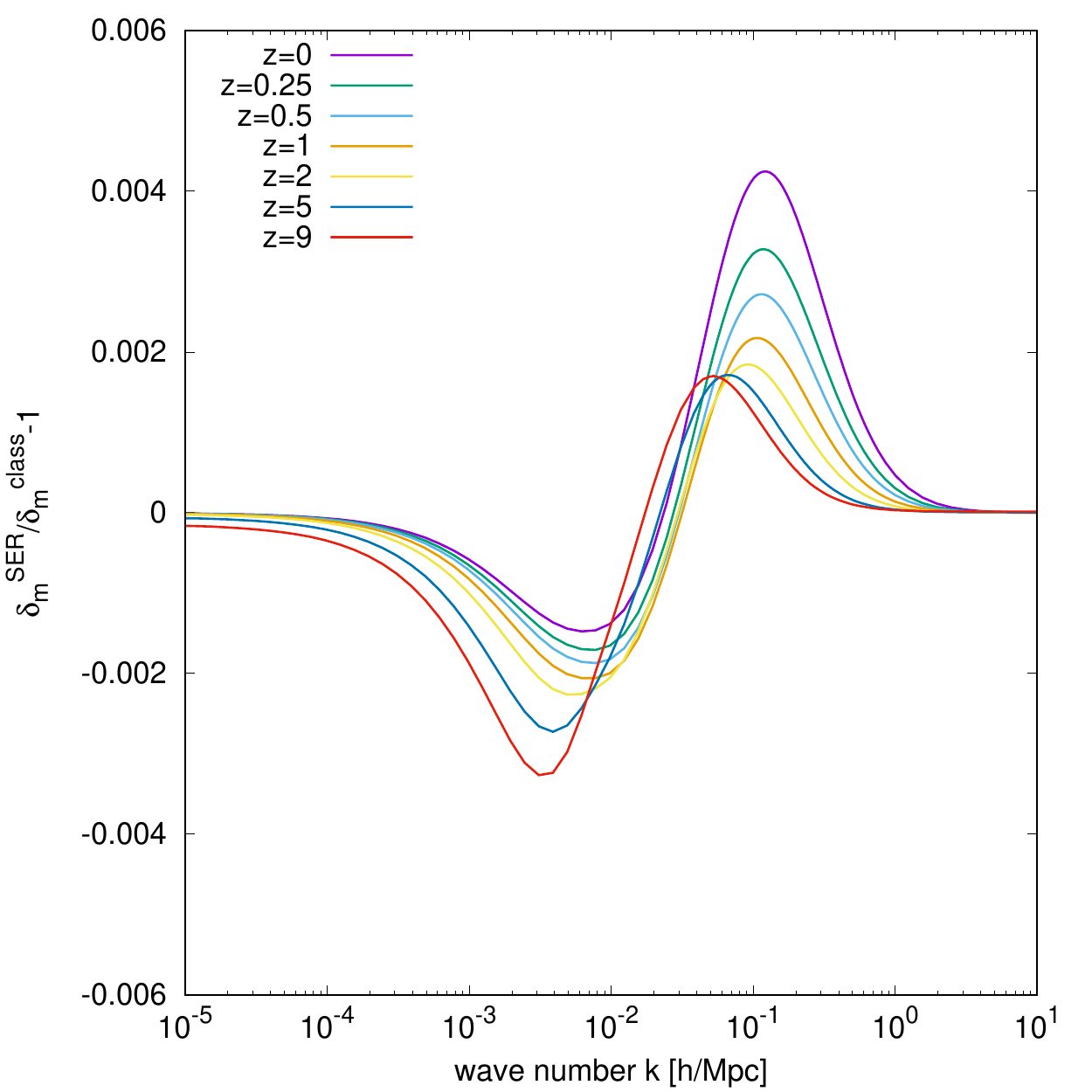}%
		\hfill
		\includegraphics[width=75mm]{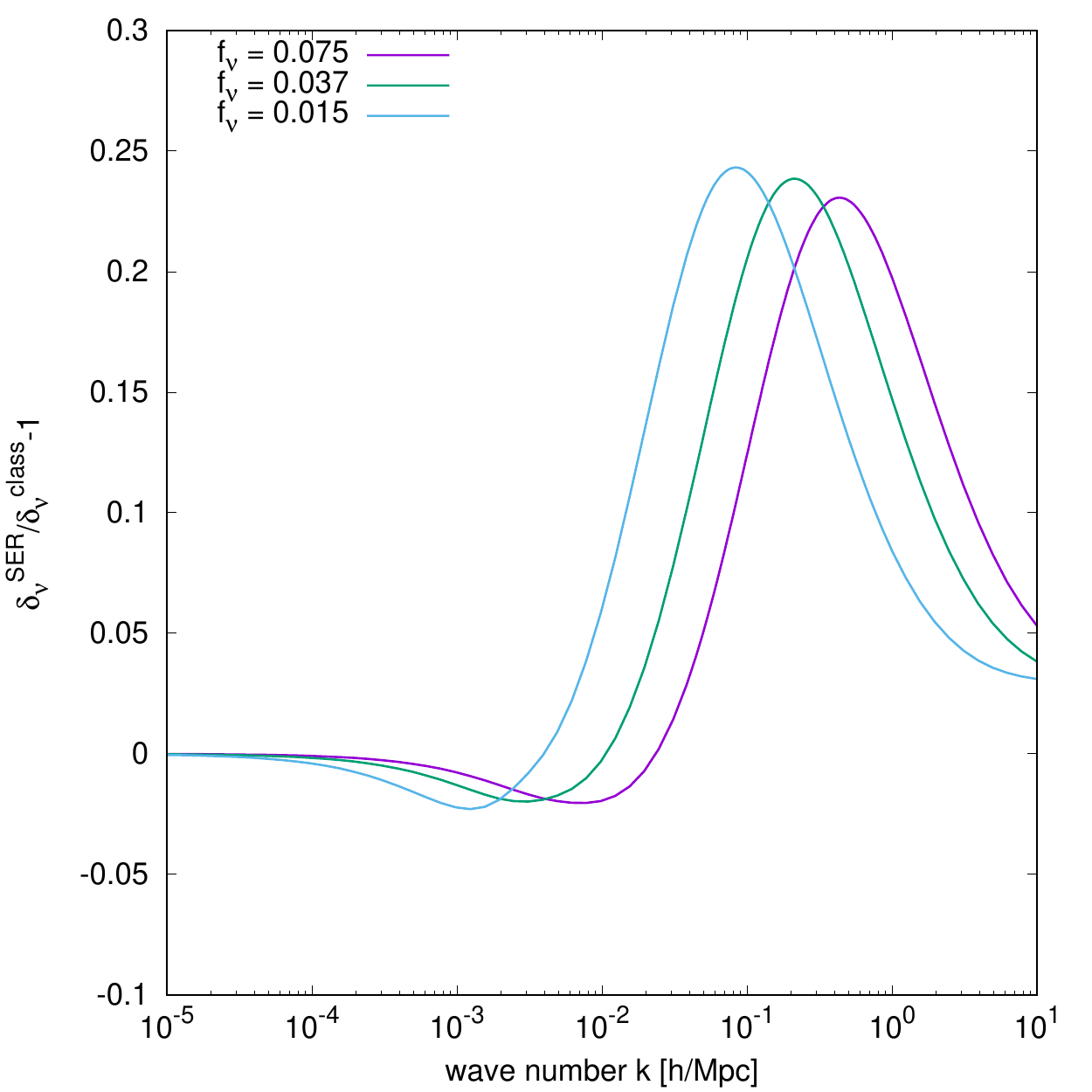}
		\includegraphics[width=75mm]{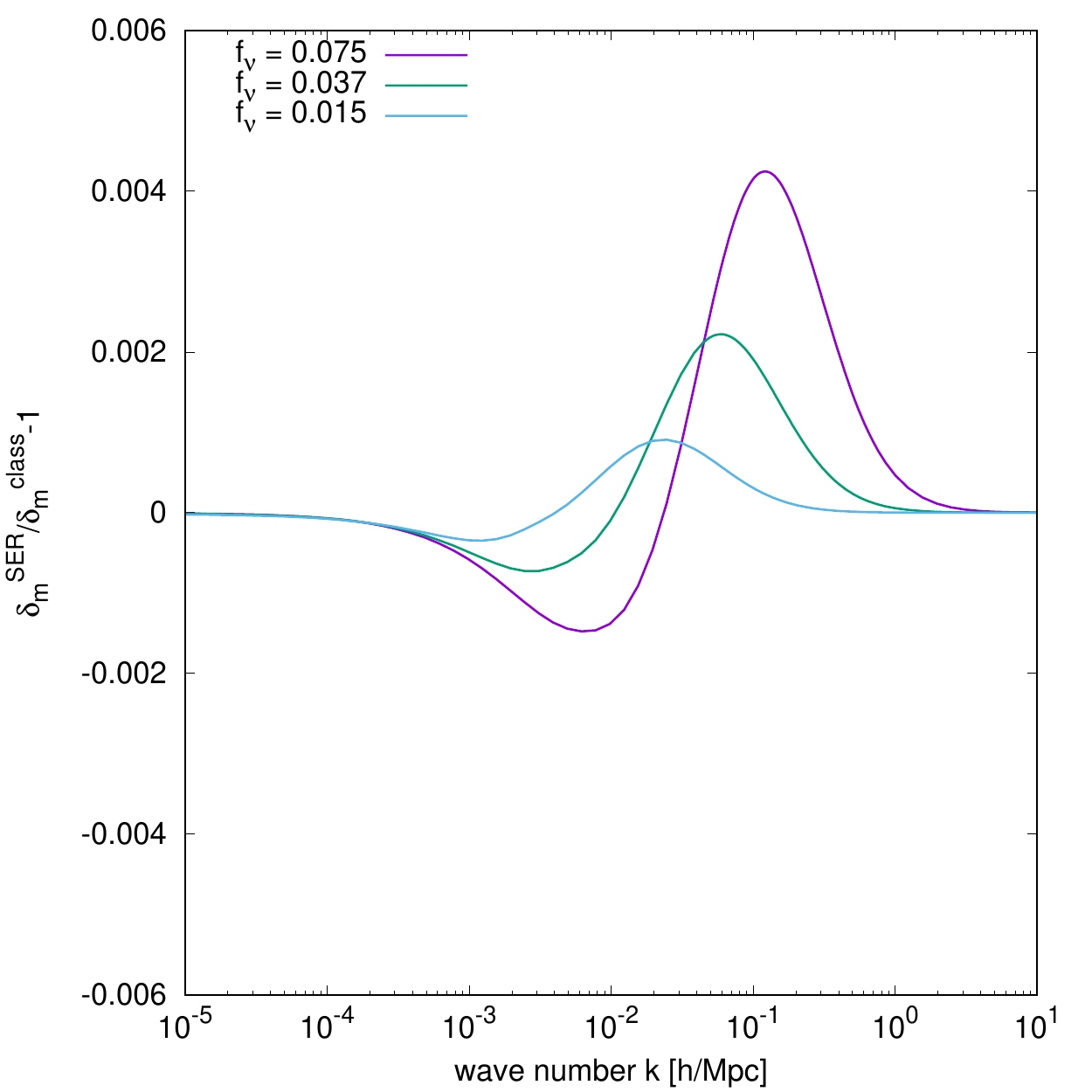}%
	\end{center}
	\caption{{\it Top left}:~Neutrino density contrast $\delta_{\nu}^{\rm SER}$ as predicted by applying the SuperEasy linear response function~\eqref{Eq:NeutrinoDensityContrastResult} to the $\delta_{\rm cb}$ output of \classcode{}, versus the actual neutrino density contrast output of~\classcode{}, $\delta_{\nu}^{\rm class}$, for the  cosmological model  Nu1  at different redshifts. 
	{\it Bottom left}:~Same as top left, but for three different values of $f_{\nu}$ corresponding to the cosmological models Nu1, Nu2, and Nu3, at $z=0$. 	
	{\it Top right}:~Same as top left, but this time comparing the total matter density contrast~$\delta_{\rm m}^{\rm SER}$ constructed from equation~\eqref{Eq:MatterDensityContrastResult} and the actual total matter density contrast output of~\classcode{}, $\delta_{\rm m}^{\rm class}$.
		{\it Bottom right}:~Same as bottom left, but for the total matter density.}
	\label{Fig:CLASSFitCheck}
\end{figure}

The integral linear response function~\eqref{Eq:MasterNeutrinoEquation} has no closed-form solution in the transition regime  between the clustering and the free-streaming limits.  However, given its limiting behaviours~\eqref{Eq:FreeStreamDeltaNuSolution}
and~\eqref{Eq:ClusteringDeltaNuSolution} at, respectively, $k \gg k_{\rm FS}$ and $k \ll k_{\rm FS}$, references~\cite{Ringwald:2004np, Wong:2008ws}  proposed an interpolation function connecting the two solutions of the form
\begin{equation}
{\delta}_{\nu} (\vec{k}, s) =\frac{k_{\rm FS}^2(s)}{[k+k_{\rm FS}(s)]^2} {\delta}_{\rm m} (\vec{k}, s),
\label{eq:interpolatem}
\end{equation}
or, equivalently,
\begin{equation}
{\delta}_{\nu}(\vec{k}, s) = \frac{k^2_{\text{FS}}(s) (1- f_{\nu})}{[k + k_{\text{FS}}(s)]^2 - k^2_{\text{FS}}(s) f_{\nu}} \, {\delta}_{\text{cb}}(\vec{k}, s) \, .
\label{Eq:NeutrinoDensityContrastResult}
\end{equation}
Then, combining equation~\eqref{Eq:NeutrinoDensityContrastResult} with the cold matter density contrast as per equation~\eqref{eq:totalcontrast} yields an interpolated response function for the total matter density contrast,
\begin{equation}
{\delta}_{\rm m}(\vec{k}, s) = \frac{[k + k_{\text{FS}}(s)]^2 (1-f_{\nu})}{[k + k_{\text{FS}}(s)]^2 - k^2_{\text{FS}}(s) f_{\nu}} \, {\delta}_{\text{cb}}(\vec{k}, s) \, .
\label{Eq:MatterDensityContrastResult}
\end{equation}
Figure~\ref{Fig:CLASSFitCheck} shows equations~\eqref{Eq:NeutrinoDensityContrastResult} and~\eqref{Eq:MatterDensityContrastResult} applied to the linear ${\delta}_{\rm cb}$ output of~\classcode{} at a range of redshifts for several massive neutrino cosmologies specified in table~\ref{Table:M000nu1Params}.

Relative to the output of~\classcode{}, the left two panels of figure~\ref{Fig:CLASSFitCheck} demonstrate that in all cases, equation~\eqref{Eq:NeutrinoDensityContrastResult} is able to reproduce the neutrino density contrast to better than 5\% accuracy in the $k \ll k_{\rm FS}$ and $k \gg k_{\rm FS}$ limits.  The interpolation is however clearly imperfect on the intermediate  scales around $k \sim k_{\rm FS}$:  here, the interpolated ${\delta}_\nu$ can deviate from the \classcode{} prediction by as much as~$\sim 20\%$ at $z=0$.  Furthermore, 
because $k_{\rm FS}$ is itself a function of redshift and the neutrino mass, the peak of the interpolation error shifts with it towards higher~$k$ as we lower the redshift and/or increase the  neutrino mass.

On the other hand, 
 because the neutrino density contrast enters the {\it total} matter density contrast along with a suppression factor $f_{\nu} \ll 1$, 
the relative difference between the interpolated ${\delta}_{\rm m}$ and its \classcode{} counterpart is, as shown in the right panels of figure~\ref{Fig:CLASSFitCheck},  always sub-percent across the whole $k$-range even for $f_\nu$ as large as $0.075$ (corresponding to $\sum m_\nu = 0.93$~eV) and improves with smaller neutrino fractions.

\begin{table}[t]
	\centering
	\begin{tabular}{c |c c c c c c  | c c c}
		\hline
		\hline 
		Model & \(\omega_{\rm m}\) & \(\omega_{\rm b}\) & \(\omega_{\nu}\) & \(h\)  & \(n_s\) & \(A_s (10^{-9})\)& $f_\nu$ & $\sum m_\nu$(eV) & $\sigma_8$\\
		\hline
		Ref1 & 0.1335 & 0.02258 & 0 &  0.71  & 0.963 & 4.0967 & 0 & 0 & 1.11 \\
		Nu1 & 0.1335 & 0.02258 & 0.01 &  0.71 & 0.963  & 4.0967 & 0.075 & 0.93 & 0.80 \\
		Nu2 & 0.1335 & 0.02258 & 0.005 &  0.71 & 0.963 &  2.9823 & 0.038 & 0.465 & 0.80 \\
		Nu3 & 0.1335 & 0.02258 & 0.002 & 0.71 & 0.963 & 2.4375 & 0.015 & 0.186 &  0.80 \\
		Nu4 & 0.1432 & 0.022 & 0.00064 & 0.67 & 0.96 & 2.1 & 0.0044 & 0.0587 & 0.81 \\
		Nu5 & 0.1432 & 0.022 & 0.00171 & 0.67 & 0.96 & 2.1 & 0.0119 & 0.1587 & 0.79 \\
		\hline
		\hline
	\end{tabular}
	\caption{Cosmological models considered in this work. We use as {\it base parameters} the present-day total matter density~$\omega_{\rm m}\equiv \Omega_{\rm m,0}h^2$, the baryon density~$\omega_{\rm b}$, the neutrino density~$\omega_{\nu}$, the dimensionless Hubble parameter~$h$, and
	the primordial scalar spectral index and amplitude $n_s$ and $A_s$.  For each model we quote also the following {\it  derived parameters}: the  neutrino fraction $f_\nu \equiv \omega_{\nu}/\omega_{\rm m}$, the neutrino mass sum $\sum m_\nu$, and $\sigma_8$, the RMS linear density fluctuation smoothed over $8/h$~Mpc.
		All models assume a flat spatial geometry, i.e., $\Omega_{\rm m}+\Omega_{\Lambda}=1$, where the $\Lambda$-component is taken to be a cosmological constant, and we split the neutrino energy density $\omega_{\nu}$ amongst three equal-mass species in the models Nu1, Nu2, and Nu3.
				The models Nu4 and Nu5 are identical to the $\nu\Lambda$CDM4 and $\nu\Lambda$CDM5 cosmologies of reference~\cite{Hannestad:2020rzl}, and assume three unequal neutrino mass values.
	\label{Table:M000nu1Params}}
\end{table}


\subsection{SuperEasy gravitational potential}
\label{sec:superfastpotential}

With equation~\eqref{Eq:MatterDensityContrastResult} we have thus arrived at the centrepiece of the SuperEasy method of massive neutrino linear response:  The impact of neutrino free-streaming on the total matter density contrast ${\delta}_{\rm m}(\vec{k}, s)$ at any one Fourier mode $\vec{k}$ and at any one time~$s$ are effectively captured by the interpolated response function~\eqref{Eq:MatterDensityContrastResult}, which is a rational function of the wave number $k$, with coefficients $f_\nu$ and $k_{\rm FS}$ given by simple algebraic expressions of the total matter density $\Omega_{\rm m,0}$, the scale factor~$a$, and the neutrino mass~$m_\nu$.  The {\it only} real-time input required is the cold matter density contrast ${\delta}_{\text{cb}}(\vec{k}, s)$.

An immediate corollary is that any cold matter-only $N$-body or hydrodynamic  large-scale structure simulation code with a PM component that utilises a Fourier grid-based force solver  can be readily adapted to include the effects of massive neutrinos via a small change to the Fourier-space Poisson equation, namely,
 \begin{equation}
 \begin{aligned}
k^2  \, {\phi}(\vec{k},s) = & \, - 4 \pi G a^2(s) \, \bar{\rho}_{\rm m} (s) \left\{\frac{[k + k_{\text{FS}}(s)]^2 (1-f_{\nu})}{[k + k_{\text{FS}}(s)]^2 - k^2_{\text{FS}}(s) f_{\nu}} \right\}  {\delta}_{\text{cb}}(\vec{k}, s) \\
=& -\, \frac{3}{2} {\cal H}^2(s)\,  \Omega_{\rm m}(s)   \left\{\frac{[k + k_{\text{FS}}(s)]^2 (1-f_{\nu})}{[k + k_{\text{FS}}(s)]^2 - k^2_{\text{FS}}(s) f_{\nu}} \right\}  {\delta}_{\text{cb}}(\vec{k}, s)\, .
\label{eq:poisson2}
\end{aligned}
\end{equation}
There are no additional equations to solve, files to read, or data to store, and no post-processing is required.
The  implementation and computational overhead of the SuperEasy linear response method is therefore virtually zero, 
making it arguably the simplest and yet analytically justifiable scheme proposed thus far to capture massive neutrino effects in simulations of large-scale structure.

Before we move on to describe how we implement the SuperEasy gravitational potential~\eqref{eq:poisson2} into a specific $N$-body code, several remarks are in order.
\begin{enumerate}
\item While equations~\eqref{Eq:NeutrinoDensityContrastResult} and hence~\eqref{Eq:MatterDensityContrastResult} have been derived for a neutrino population characterised by one free-streaming scale $k_{\rm FS}$, they are easily generalisable to scenarios with multiple free-streaming lengths (due to e.g., a realistic neutrino mass hierarchy).
See appendix~\ref{Sec:ExtHierarchy}  for details.

\item That the interpolated neutrino density contrast~\eqref{Eq:NeutrinoDensityContrastResult} has a 20\% error around $k \sim k_{\rm FS}$ may be a nagging concern.  To improve upon this state of affairs one could in principle calibrate the interpolated response function to the particular cosmology under investigation on a case-by-case basis.  

This might be achieved, for example, by identifying the ratio ${\delta}_{\nu}/{\delta}_{\rm cb}$ with the corresponding ratio formed from the linear transfer function outputs of \cambcode{} or \classcode{}, and pre-compute the ratio at the Fourier grid points and redshift time steps  required by the $N$-body simulation.  A similar scheme is also sometimes applied in perturbative analyses (e.g.,~\cite{Aviles:2020cax,Garny:2020ilv}) and there are some parallels with the grid-based linear neutrino perturbation method of reference~\cite{Brandbyge:2008js} (which uses the linear $\delta_{\nu}$ output of \cambcode{} rather than the ratio ${\delta}_{\nu}/{\delta}_{\rm cb}$).

In any case, we deem the benefit of further calibration of equation~\eqref{Eq:NeutrinoDensityContrastResult} to be highly diminishing given the cost, if sub-percent accuracy is demanded ultimately only of the {\it total} matter power spectrum prediction. For this reason we do not advocate it.

\item Notwithstanding the sub-percent agreement between equation~\eqref{Eq:MatterDensityContrastResult} and the {\it linear} output of \classcode{}, it is a legitimate concern that the same may not materialise once nonlinear cold matter dynamics have been incorporated into the calculation.  

To this end, we recall that the derivation of the free-streaming solution~\eqref{Eq:FreeStreamDeltaNuSolution} makes no assumption about the linearity or otherwise of the external  source~${\phi}(\vec{k},s)$ to which the neutrinos respond.  The clustering solution~\eqref{Eq:ClusteringDeltaNuSolution}, on the other hand, belongs in the inherently linear regime where the \classcode{} outputs apply.
 Thus, if the interpolation function~\eqref{Eq:MatterDensityContrastResult} should perform more poorly in the presence of nonlinear 
cold matter dynamics,  we would expect any deviation from the full integral linear response~\eqref{Eq:MasterNeutrinoEquation} to show up most prominently around $k \sim k_{\rm FS}$.  

We can test this understanding and hence the validity of the SuperEasy method by direct comparison with an explicit numerical evaluation of the integral linear response function~\eqref{Eq:MasterNeutrinoEquation} within an $N$-body code in the manner of~\cite{AliHaimoud:2012vj}.  See section~\ref{Sec:ComparisonILR}.
 
\item  Ultimately, what makes linear response --- even the integral version~\eqref{Eq:MasterNeutrinoEquation} --- numerically relatively easily  tractable is linearisation of the collisionless Boltzmann equation, i.e., the approximation~\eqref{eq:linearisation}. But the assumption of linearity must break down for sufficiently large neutrino masses, and quantifying the validity region of linear response in terms of how well the approximation~\eqref{eq:linearisation} and the accompanying linearisation condition~\eqref{eq:linearisationcondition} are satisfied must also form part of the investigation of such methods.  We shall discuss this point in detail in 
section~\ref{subsec:neutrino_density_contrast_from_linear_response}.

\end{enumerate}
Then, without further ado, we now proceed to describe how to incorporate the SuperEasy massive neutrino linear response method in an $N$-body simulation code.


\section{SuperEasy linear response in $N$-body simulations}
\label{Sec:Nbody}

We demonstrate our SuperEasy linear response method by implementing it into the publicly available $N$-body simulation code~\gadgetcode{} \cite{Springel:2005mi}. The specific modifications to the code and the initial condition generator~\ngenic{}~\cite{Angulo:2012ep} are outlined in two subsections below.  We stress however that the implementation does not in any way depend on, or is specifically designed for, the architecture of \gadgetcode{} or \ngenic{}.
  Indeed, the power of the SuperEasy method lies in its simplicity, and what we present below is largely intended as a recipe applicable to any $N$-body code containing a PM component.


\subsection{Modified \gadgetcode{}}
\label{Sec:Gadget}

The stock version of \gadgetcode{} is a hybrid PM code with an additional oct-tree based short-range force solver. In our modifications, the interface with massive neutrino linear response happens solely on the mesh (or ``grid'') within the PM component; the portion of the code that  governs directly the particle dynamics remains  unchanged from the stock version.

The justification for this choice of modification stems from the fact the PM-grid is generally finer than the free-streaming scale of neutrinos, so that any sub-grid tree-level force within the neutrino sector is subdominant for convergence considerations. As an illustration, the simulations presented in this work use PM-grids with a maximum cell length of $0.5 \, \text{Mpc}/h$. In comparison, the free-streaming scale of three degenerate neutrinos with masses summing to $\sum m_{\nu} = 0.93 \, \text{eV}$  is approximately $26 \, \text{Mpc}/h$ at $z=0$.  At higher redshifts, or for lower masses, the free-streaming scale will be even larger than this estimate. We therefore conclude that neutrino clustering can be modelled with sufficient accuracy using only the PM component.

 In a standard cold matter-only simulation, the Newtonian gravitational force in the PM component is calculated at every time step first by distributing the simulation particles onto the PM-grid points using an interpolation scheme such as the Cloud-in-Cell (CIC) method; this gives us $\delta_{\rm cb}(\vec{x},s)$.  A discrete Fourier transform turns $\delta_{\rm cb}(\vec{x},s)$ into ${\delta}_{\rm cb}(\vec{k},s)$, from which we can solve for the corresponding $k$-space potential ${\phi}(\vec{k},s)$ and hence forces on the Fourier-grid points via the Poisson equation.  A subsequent inverse Fourier transform brings us back to coordinate space, where the real-space forces on the PM-grid points can now be interpolated to the particle positions, through which to forward the particle trajectories.

The entry point for incorporating the SuperEasy massive neutrino linear response method into the PM component is the Fourier grid.  Here, we replace the standard Poisson equation with the interpolated one~\eqref{eq:poisson2}, where ${\delta}_{\rm cb}(\vec{k},s)$ is identified with the Fourier transform of the cold matter density contrast~$\delta_{\rm cb}(\vec{x},s)$ constructed from smoothed particles on the PM-grid points.  Then, subsequent standard evaluations of the gravitational force on the simulation particles will automatically account for the effects of massive neutrinos.

One further though optional point of modification is the Hubble expansion rate, which, depending on the initialisation time of the simulation, may need to be corrected for a small deviation from the $\bar{\rho}_\nu \propto a^{-3}$ behaviour in the neutrino sector due to the relativistic to nonrelativistic transition.  In practice, however, this correction has no discernible effect on the final outcome, {\it as long as}  the same Hubble expansion rate is used in both the $N$-body code and in the initialisation procedure.  See  below.


\subsection{Initial conditions}
\label{Sec:IC}

Because the actual simulation uses a particle realisation only for the cold matter species, the initial conditions for the simulation particles can be generated  in essentially same way as a conventional cold matter-only simulation. 

We adopt the  standard ``scale-back'' method when setting the initial conditions of our simulations, which circumvents the need to correct for relativistic effects within the $N$-body code itself.   In this approach, the linear cb power spectrum of the cosmology of interest is first calculated with a fully relativistic Boltzmann code such as~\classcode{} to $z=0$, and then scaled back to the simulation initial redshift~$z_{\rm i}$ using a set of ``fictitious'' linear growth factors~$D_{\rm cb}(k, a)$ computed under the assumption of Newtonian gravity.  This ensures that the simulation outcome will match the fully relativistic linear theory predictions at $z=0$ on large scales, with the simulation itself serving only to produce nonlinear enhancements on small scales.

Our fictitious linear growth factors are  calculated from the linearised continuity and Euler equations (see, e.g., \cite{Bernardeau:2001qr}), together with the SuperEasy Poisson equation~\eqref{eq:poisson2} incorporating the influence of the neutrino density contrast~$ \delta_\nu$.  In Fourier space and using the scale factor $a$ as a time variable, these equations take the form
\begin{equation}
	\begin{aligned}
		\frac{\mathrm{d}D_{\text{cb}}(k, a)}{\mathrm{d} a} = & \, \theta_{\text{cb}}(k, a), \\
			\frac{\mathrm{d}\theta_{\text{cb}}(k, a)}{\mathrm{d} a} = & \, - \left( 2 + \frac{\mathrm{d} \ln \mathcal{H}}{\mathrm{d} \ln a} \right) \frac{1}{a} \theta_{\text{cb}}(k, a) - \frac{k^2 \phi}{a^2 {\cal H}^2} ,
		\label{eq:dcbeom}
	\end{aligned}
\end{equation}
where in the SuperEasy method the gravitational potential $\phi$ is, as mentioned above, given by equation~\eqref{eq:poisson2}.
Importantly, because neutrino free-streaming introduces a scale-dependence, the linear growth factors obtained from solving this set of equations are necessarily $k$-dependent, and we stress again that the conformal Hubble expansion rate ${\cal H}$ used  to solve equation~\eqref{eq:dcbeom} must match that implemented in the $N$-body code itself.

The cold matter particle realisation is performed via the Zel'dovich approximation (ZA) at $z_{\rm i}= 49$ using a modified version of the publicly available \ngenic{} code~\cite{Angulo:2012ep}.   Modification is necessary because  the stock version of \ngenic{} code computes the initial particle velocities entirely in {\it real space} via a simple product 
\begin{equation}
	\vec{u}(\vec{q}, a) = - \mathcal{H}(a) \, \frac{\mathrm{d}\ln{D_{\text{cb}}(q,a)}}{\mathrm{d}\ln{a}} \, \vec{\Psi}(\vec{q}, a) \, ,
	\label{eq:realvel}
\end{equation}
where $\vec{\Psi}(\vec{q}, a)$ is the real-space random displacement field consistent with the input initial cb power spectrum.  In a massive neutrino cosmology, however, because of the inherent $k$-dependence of the linear growth function $D_{\rm cb}(k, a)$, 
the simple product needs to be promoted to a convolution product,
\begin{equation}
	\vec{u}(\vec{q},a) = - \mathcal{H}(a) \, {\cal F}^{-1} \left[\frac{\mathrm{d}\ln{D_{\text{cb}}(k,a)}}{\mathrm{d}\ln{a}} \right]  * \vec{\Psi}(\vec{q}, a) \, ,
\end{equation}
where $A(x)={\cal F}^{-1}[A(k)]$ denotes an inverse Fourier transform.   In practice, this means it may be most convenient to compute first {\it both}  $\vec{\Psi}(\vec{k}, a)$ and  $\vec{u}(\vec{k}, a)$ in Fourier space, before performing the inverse Fourier transform  to real space.


\section{Convergence tests}
\label{Sec:Results}

We run simulations in this work for several massive neutrino and $\Lambda$CDM cosmologies specified by the cosmological parameter values given in table~\ref{Table:M000nu1Params}.  The non-neutrino parameters of the first four models all have values lying within 1$\sigma$ of the WMAP 9-year best-fits~\cite{Bennett:2012zja}, while those of the last two models are 1$\sigma$ compatible with the Planck 2018 best-fits. 
Of the five massive neutrino cosmologies, only Nu1 has a neutrino mass sum exceeding even the most conservative of the current $2 \sigma$~observational limits~\cite{Oldengott:2019lke}. 
Because it is the most ``extreme'' cosmology under consideration, we also base our convergence tests on the Nu1 model.

We use a fixed simulation volume of $V=512^3 \, (\text{Mpc}/h)^3$, so that the smallest wave number ($k \sim 0.01 h/\text{Mpc}$) captured falls well within the linear regime to justify the scale-back initialisation method (see section~\ref{Sec:IC}).  Each simulation uses the same PM-grid with a cell length of $0.5 \, \text{Mpc}/h$. The same grid is also used for power spectrum estimation.   We vary the number of cold matter simulation particles  between  $N=128^3$ and $N=1024^3$ to test for convergence, and adjust the force-softening length in the tree component accordingly to one-twentieth of the average interparticle spacing in each case.


\subsection{Absolute power spectrum}

\begin{figure}[t]
	\begin{center}
		\includegraphics[width=75mm]{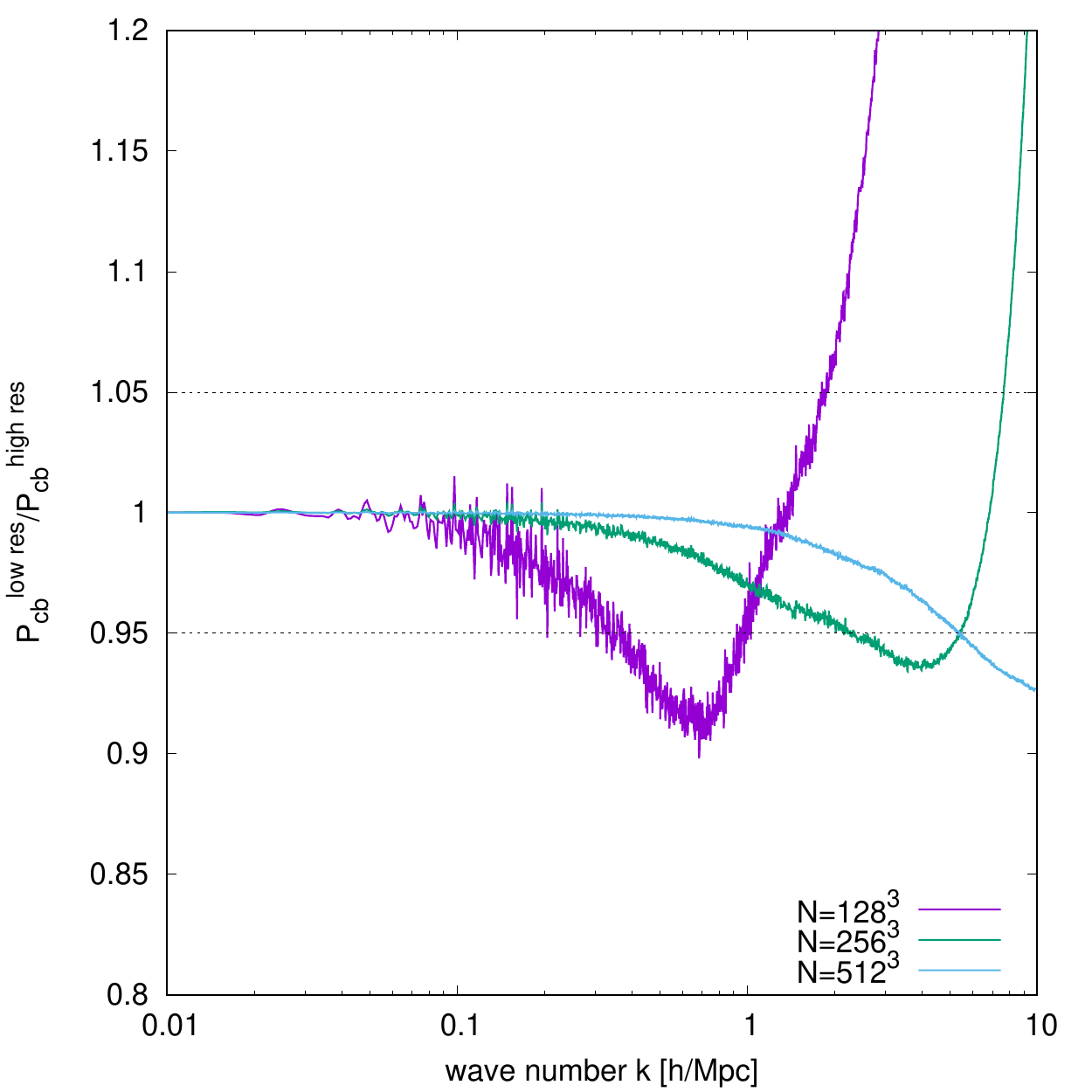}
		\includegraphics[width=75mm]{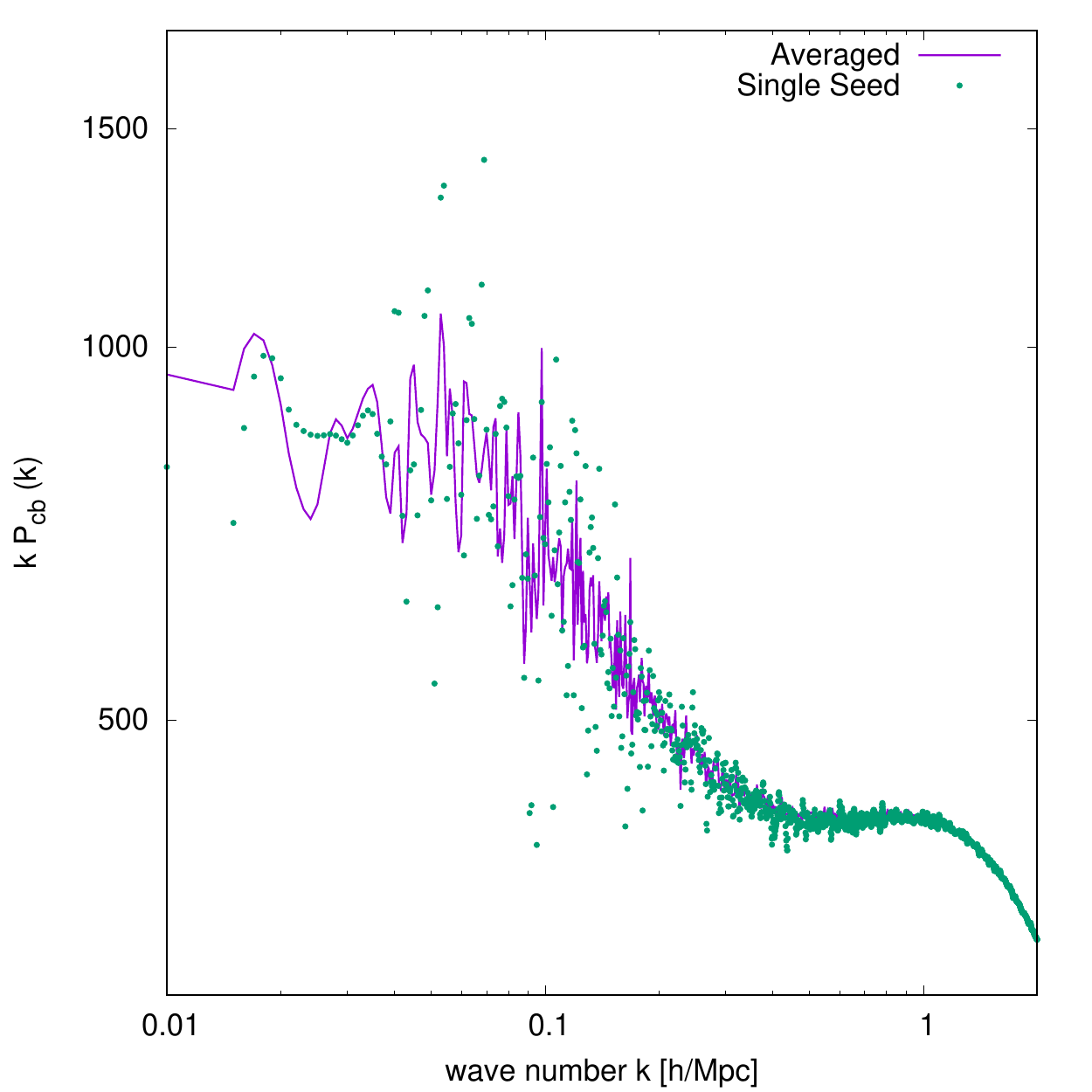}%
	\end{center}
	\caption{{\it Left}: Present-day ($z=0$) cb power spectra  extracted from simulations using $N=128^3$ (purple), $256^3$ (green), and $512^3$ (blue) particles for the cosmological model Nu1, all normalised to a high-resolution $N=1024^3$ power spectrum for the same cosmology.  Clearly, increasing the mass resolution via increasing $N$ stabilises the small-scale power prediction. 
		{\it Right}:  Power spectrum predictions at $z=0$ obtained from (i)~one single realisation (dots) and (ii) averaging over five different sets of initial seeds (solid line). The averaging procedure significantly reduces the scatter in the low-$k$ region arising from sample variance.}
	\label{Fig:ConvergenceTest}
\end{figure}

Fixing the simulation volume at $V=512^3 \, (\text{Mpc}/h)^3$ while varying the number of simulation particles~$N$, the left panel of figure~\ref{Fig:ConvergenceTest} shows the cb power spectra for the Nu1 model at $z=0$ extracted from the $N=128^3, 256^3, 512^3$ runs, normalised to  the $N=1024^3$ result --- the highest resolution achievable with our computing facilities.

On large scales, the excellent agreement between different choices of particle number~$N$ is guaranteed by the identically sized Fourier grid ($1024^3$ cells) used for the initialisation of all simulations, irrespective of particle numbers~$N$. In this way, the smaller-$N$ runs are essentially a downsampling of the initial density field onto coarser grids, a procedure that leaves the large-scale correlations (i.e., small-$k$ power) largely unaffected between different choices of particle number $N$.

The quasi-linear and nonlinear regimes, $k\gtrsim 0.1 \, h/\text{Mpc}$, however, typically become first under- and then over-powered in the low-resolution simulations. The former, under-powering phenomenon can be attributed to the finite mass resolution: the further the separation between  simulation particles and the higher the mass each particle carries, the less efficiently the particles can  cluster nonlinearly.
 The latter, over-powering effect is, on the other hand, a manifestation of Poisson noise.
 With a larger number of simulation particles, both of these artefacts can be pushed to a larger value of $k$. Evidently in the left panel of figure~\ref{Fig:ConvergenceTest},  percent-level agreement between the $N=512^3$ and $1024^3$ runs can be achieved up to  $k\sim 2 \, h/\text{Mpc}$. 
 
The right panel of figure~\ref{Fig:ConvergenceTest} compares the cb power spectrum extracted from a single realisation and one formed from averaging over five power spectra computed using five different sets of initial seeds. As expected, the low-$k$ region most susceptible to sample variance suffers from less scatter when averaged over five different sets of  seeds.  Conversely, the high-$k$ region is largely unaffected by averaging, since there is inherently enough volume here to overcome sample variance even in one single realisation.

\begin{figure}
	\centering
	\includegraphics[scale=1]
	{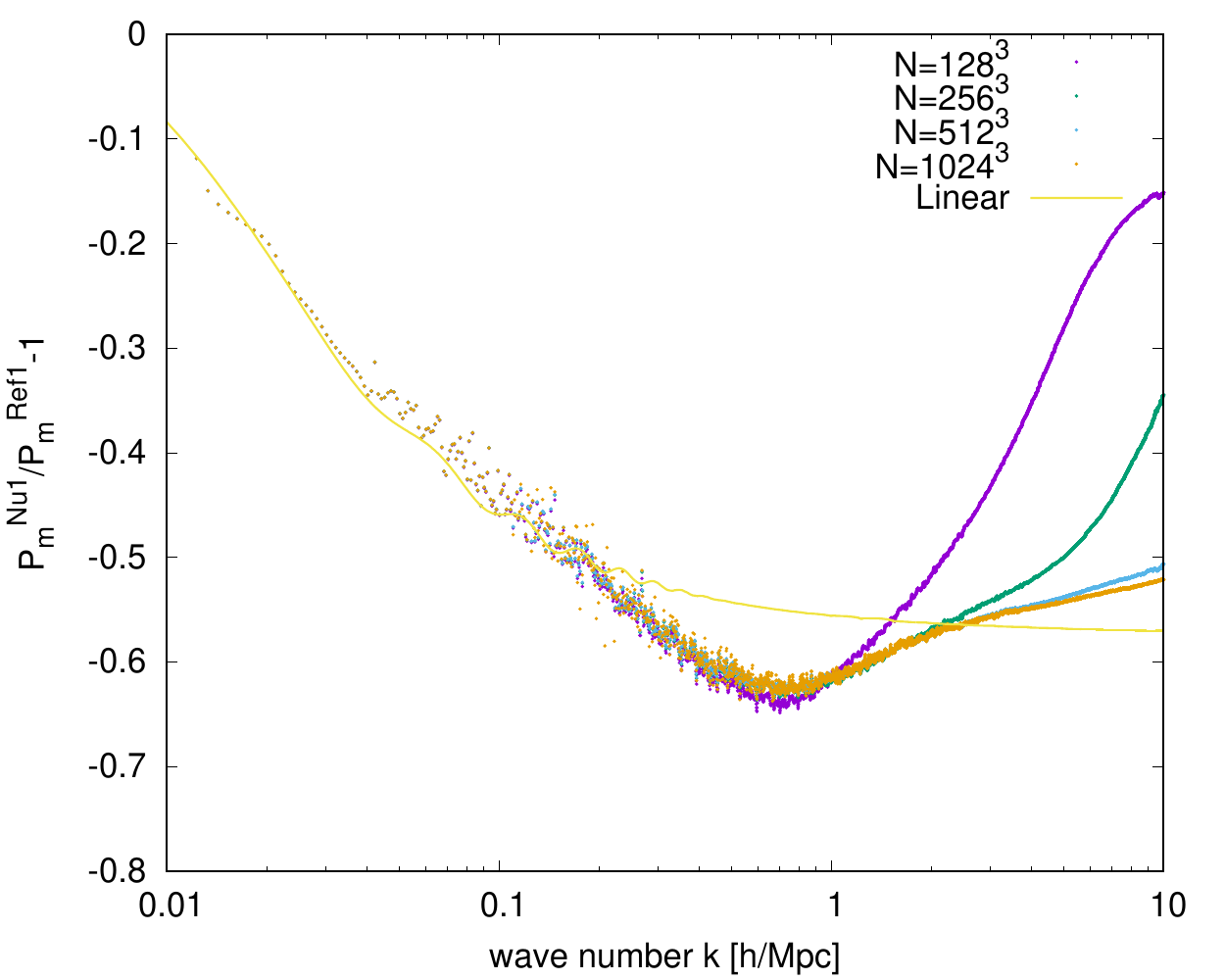}
	\caption{The relative total matter power spectra at $z=0$ between the models Nu1 and Ref1 obtained from four pairs of simulations with increasing mass resolution/particle number~$N$. All simulations have been performed in a box of volume $V=512^3 \, (\text{Mpc}/h)^3$. The solid yellow line indicates the corresponding linear theory result from~\classcode{}.
		\label{Fig:SpoonConvergence}}
\end{figure}


\subsection{Relative power spectrum}

Convergence can also be demonstrated in the  {\it relative} total matter power spectrum, defined here as the  power spectrum ratio of a massive neutrino cosmology to the reference $\Lambda$CDM model simulated under identical conditions, including identical  initial seeds.
  Figure~\ref{Fig:SpoonConvergence} shows four sets of relative total matter power spectra between the models Nu1 and Ref1, computed using four choices of particle numbers ($N=128^3, 256^3, 512^3, 1024^3$).  
  
At  $k\lesssim 0.1 \, h/\text{Mpc}$, the relative power is consistent across all mass resolutions tested.   For all four choices of $N$ we observe a spoon-like dip followed by an upturn feature, consistent with previous findings~\cite{Brandbyge:2008rv, Bird:2011rb, Bird:2018all, Partmann:2020qzb}.  The physical origin and nature of the spoon feature have been discussed in detail in terms of the halo model~\cite{Hannestad:2020rzl}.
Here, we merely observe that the dip occurs at the same $k \sim 0.7\, h/{\rm Mpc}$ in all runs; the low-resolution runs tend however to predict too deep a dip followed by too high an upturn at $k\gtrsim 1 \, h/\text{Mpc}$.  As in figure~\ref{Fig:ConvergenceTest}, these low-resolution artefacts seen in figure~\ref{Fig:SpoonConvergence} are
 a consequence first of power underestimation due to poor mass resolution and then of high-$k$ Poisson noise, both of which dominate over  neutrino free-streaming effects on small scales.

On the other hand, the higher-resolution ($N=512^3, 1024^3$)  relative power spectrum predictions exhibit sub-percent level agreement up to $k \sim 6 \, h/\text{Mpc}$, exceeding the $k\sim 2 \, h/\text{Mpc}$ limit on the concordant region seen in figure~\ref{Fig:ConvergenceTest} for their absolute power spectrum counterparts.
  This observation is consistent with the proposition of~\cite{Hannestad:2019piu}, which states that the relative power spectrum generally exhibits better convergence properties than can be achieved for the absolute power spectrum using the same simulation settings.

Henceforth, unless otherwise specified, we shall always use the highest-resolution setting ($N=1024^3, \, V=512^3 \, (\text{Mpc}/h)^3$) achievable on our computing facilities in our simulations.  Where comparisons are called for (e.g., between cosmological models, between methods, etc.), we always compare simulation results initialised with the same seeds.


\section{Comparison with other approaches}
\label{sec:comparison}

Having identified a suitable choice of simulation settings,  we are now in a position to contrast the SuperEasy nonlinear power spectra with predictions from our own implementation of the integral linear  response method~\cite{AliHaimoud:2012vj} in \gadgetcode{}, as well as with the results of reference~\cite{Hannestad:2020rzl} obtained from grid-based linear neutrino simulations.  The latter comparison is especially interesting in that, not only is the grid-based linear neutrino method a different model of neutrino perturbations in $N$-body simulations compared with linear response, reference~\cite{Hannestad:2020rzl} also used a different simulation code, 
\pkdgrav{}~\cite{Potter:2016ttn}.


\subsection{Comparison with integral linear response}
\label{Sec:ComparisonILR}

Let us compare first  our SuperEasy  method with the original integral linear response approach of reference~\cite{AliHaimoud:2012vj}, which explicitly solves the integral linear response function~\eqref{Eq:MasterNeutrinoEquation}  within the $N$-body code to obtain the neutrino density contrast at every time step.  Other than this integral evaluation, the two methods share an identical perturbative starting point for the neutrino sector and evolve the cold sector in the presence of neutrino clustering in exactly the same way.
The comparison is therefore largely  straightforward, save for some details in the implementation of the response function~\eqref{Eq:MasterNeutrinoEquation}  in \gadgetcode{}, to be discussed below.

\begin{figure}[t]
	\centering
	\includegraphics[scale=1]
	{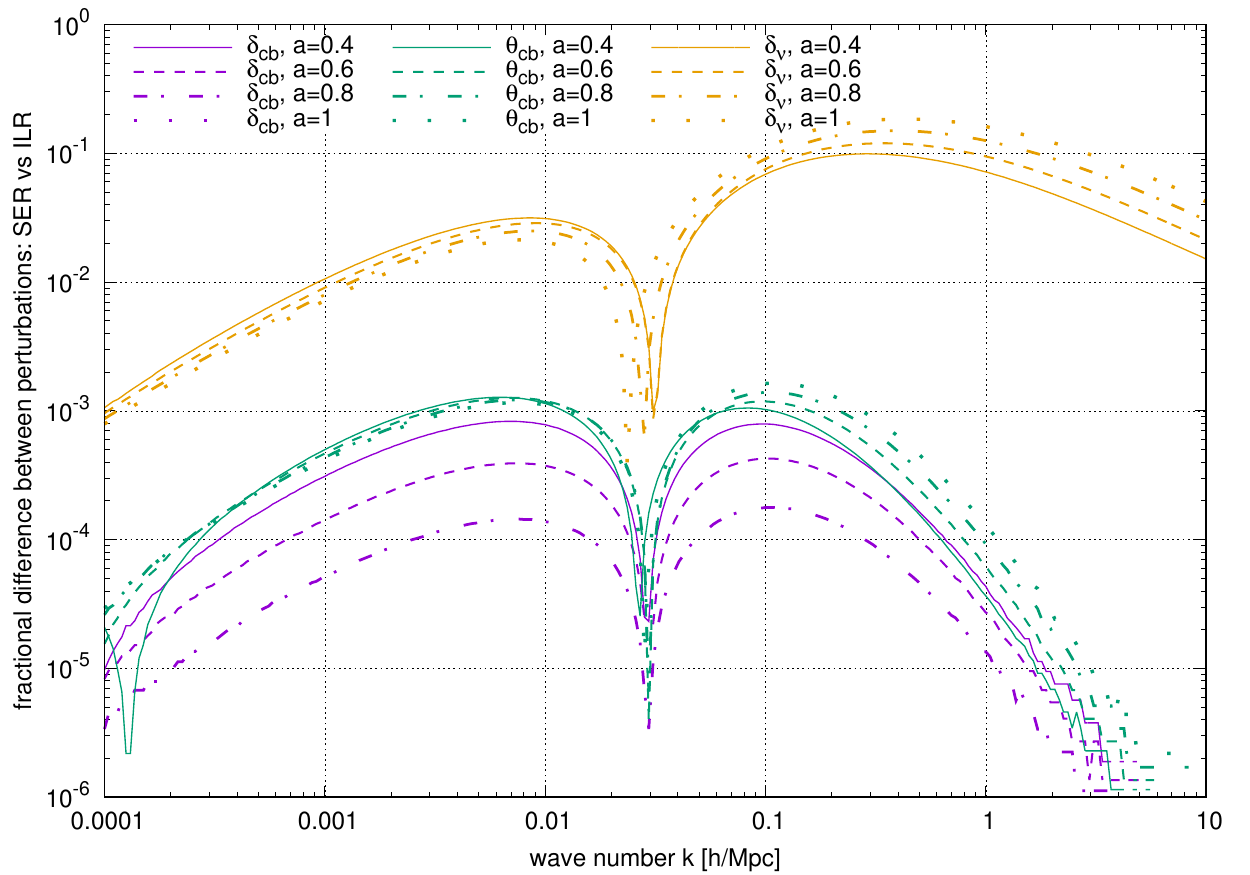}
	\caption{Fractional differences between the predictions of SuperEasy and integral linear response, applied to a {\it linearly} evolved cb density contrast, for the model Nu1. Agreement between the cb density contrasts $\delta_{\rm cb}$ improves with lower redshifts, while the differences in $\theta_{\text{cb}}$ and in $\delta_{\nu}$ remain fairly stable with time. The largest discrepancy in $\delta_{\nu}$ is consistent with the results of figure \ref{Fig:CLASSFitCheck}.}
	\label{Fig:FLR_ILR_Res}
\end{figure}

As a warm-up exercise, we first contrast the two solutions to the {\it linearised} equation of motion~\eqref{eq:dcbeom} for the cb density contrast 
 ${\delta}_{\rm cb}(k)$ and velocity divergence~${\theta}_{\rm cb}(k)$, 
 using in the gravitational potential~$\phi$ the neutrino density contrast ${\delta}_\nu(k)$ computed from (i)~the SuperEasy response function~\eqref{Eq:NeutrinoDensityContrastResult}, and (ii)~the full integral response function~\eqref{Eq:MasterNeutrinoEquation}.  As in the scale-back initialisation procedure, we normalise  ${\delta}_{\rm cb}$ at  $z=0$. 
 Figure~\ref{Fig:FLR_ILR_Res} shows the fractional differences  in ${\delta}_{\rm cb}$, ${\theta}_{\rm cb}$, and ${\delta}_{\nu}$ between these approximations for the model Nu1 at a range of redshifts.  
 Clearly, all three quantities exhibit two peaks in the intermediate $k$-region,  indicating the same interpolation discrepancy previously observed in figure~\ref{Fig:CLASSFitCheck} in our comparison of the SuperEasy and  \classcode{} outputs.  The maximum discrepancy in ${\delta}_\nu$ is again about 20\%, while the differences in ${\delta}_{\rm cb}$ and in ${\theta}_{\rm cb}$ are at most 0.2\%.  The agreement in ${\delta}_{\text{cb}}$ improves with lower redshift, but this merely reflects our choice of normalisation at $z=0$.

Next, we compare the nonlinear outputs of the two methods embedded in an $N$-body simulation.  To this end, we implement the integral linear response function~\eqref{Eq:MasterNeutrinoEquation} in \gadgetcode{} following reference~\cite{AliHaimoud:2012vj}.  As with  SuperEasy linear response, the entry point for the implementation of integral linear response  is the Fourier grid and hence the Poisson equation in the PM component.  However, because the integral linear response function~\eqref{Eq:MasterNeutrinoEquation}  involves a time-integration of the gravitational potential evolution history on each $\vec{k}$-grid point, to save computing resources reference~\cite{AliHaimoud:2012vj} advocates (i)~storing only one  $\langle {\phi} (\vec{k})\rangle$ per $k \equiv |\vec{k}|$ mode averaged  over all $\hat{k}$ directions, i.e., identical $|{\delta}_\nu(\vec{k})|$ for all vectors $\vec{k}$ with the same magnitude~$k$, and (ii)~the final ${\delta}_\nu(\vec{k})$  on a $\vec{k}$-grid point assumes the same complex phase as the cb density contrast ${\delta}_{\rm cb}(\vec{k})$  on that same $\vec{k}$-grid point.  To make the comparison as direct as possible, we  implement the same averaging procedure in our SuperEasy simulations, although in practice this extra step makes no more than a 0.1\% difference to the final result, and, from the SuperEasy perspective, in fact detracts from the simplicity of the method.

\begin{figure}[t]
	\centering
	\includegraphics[scale=1]
	{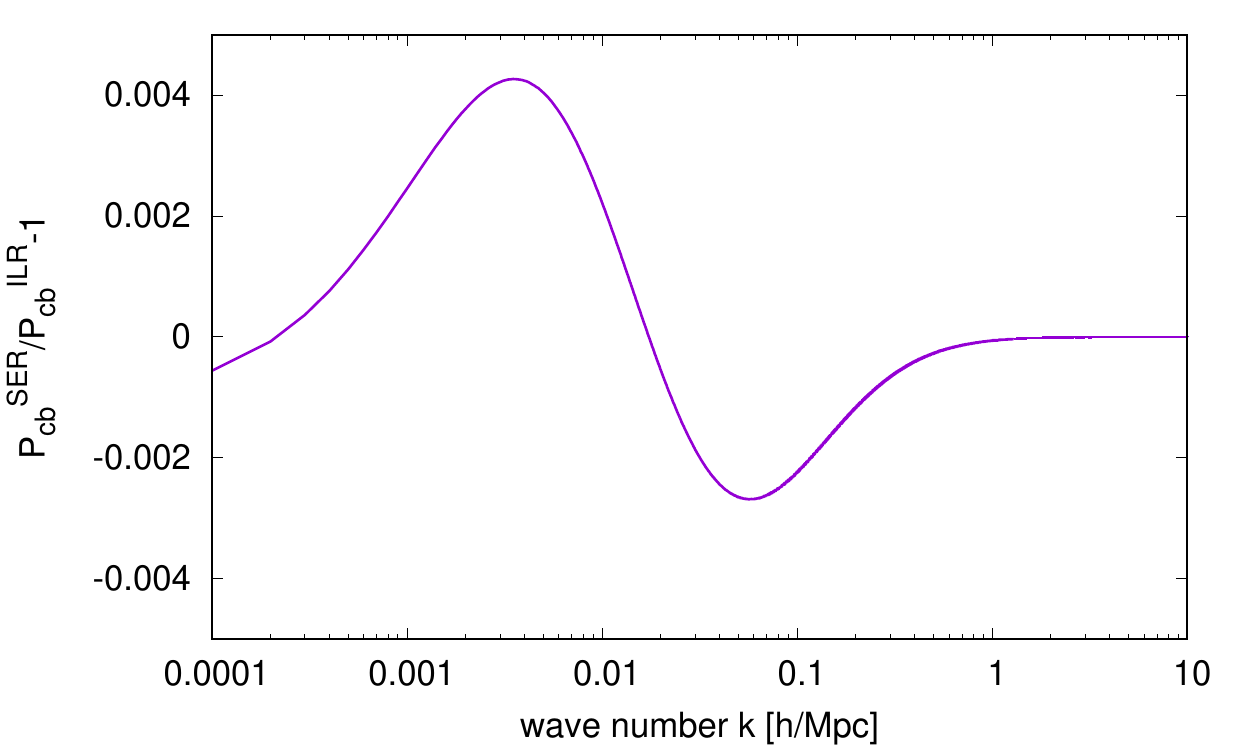}
	\caption{Fractional difference in the initial ($z=49$) cb power spectra of the Nu1 model, constructed using the scale-back method for a SuperEasy simulation, $P_{\rm cb}^{\rm SER}$,  and for an integral linear response simulation, $P_{\rm cb}^{\rm ILR}$.  The largest difference, 0.4\%, occurs at a fairly small $k \sim 0.003\, h/{\rm Mpc}$.  This difference disappears as we forward the systems in time, because of the matching condition imposed on the small-$k$ power spectrum that forms part of the scale-back initialisation procedure.}
	\label{Fig:ICComparison}
\end{figure}

Likewise, the Hubble expansion history and  time step divisions in the execution of the simulations are kept consistent across the comparison.  Initialisation of the integral linear response simulation follows the same scale-back procedure described in section~\ref{Sec:IC}, but with the neutrino density contrast~${\delta}_\nu(k)$  now computed from the integral linear response function~\eqref{Eq:MasterNeutrinoEquation} instead of the SuperEasy response function~\eqref{Eq:NeutrinoDensityContrastResult}.
  Figure~\ref{Fig:ICComparison} shows the fractional difference between the two $z=49$ cb power spectra used to initialise the SuperEasy and the integral response linear simulations of the Nu1 model.   The biggest difference shows up on the  large scales  around $k \sim 0.003~h/{\rm Mpc}$, but still remains within 0.4\% agreement.

\begin{figure}[t]
	\begin{center}
		\includegraphics[width=75mm]{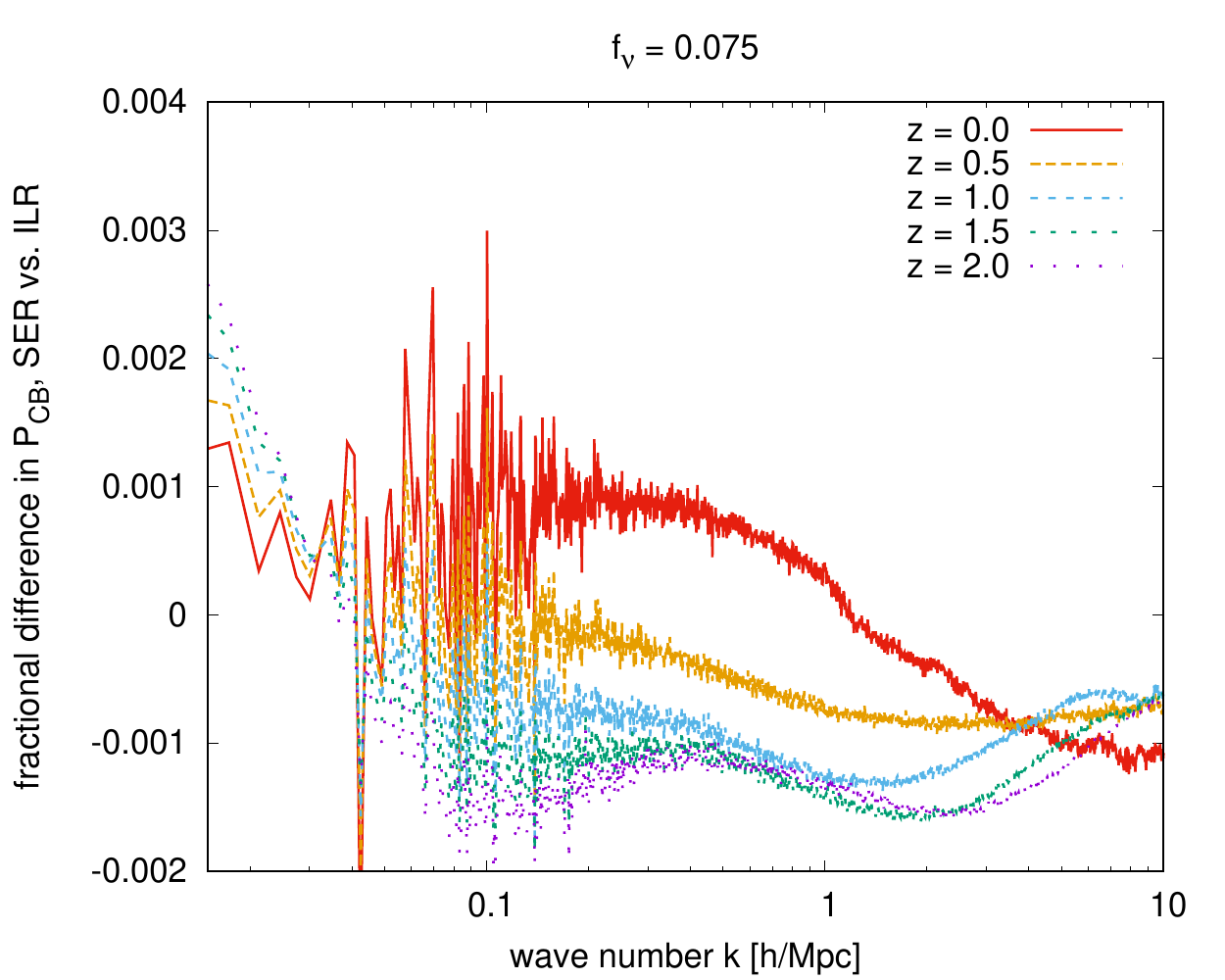}
		\includegraphics[width=75mm]{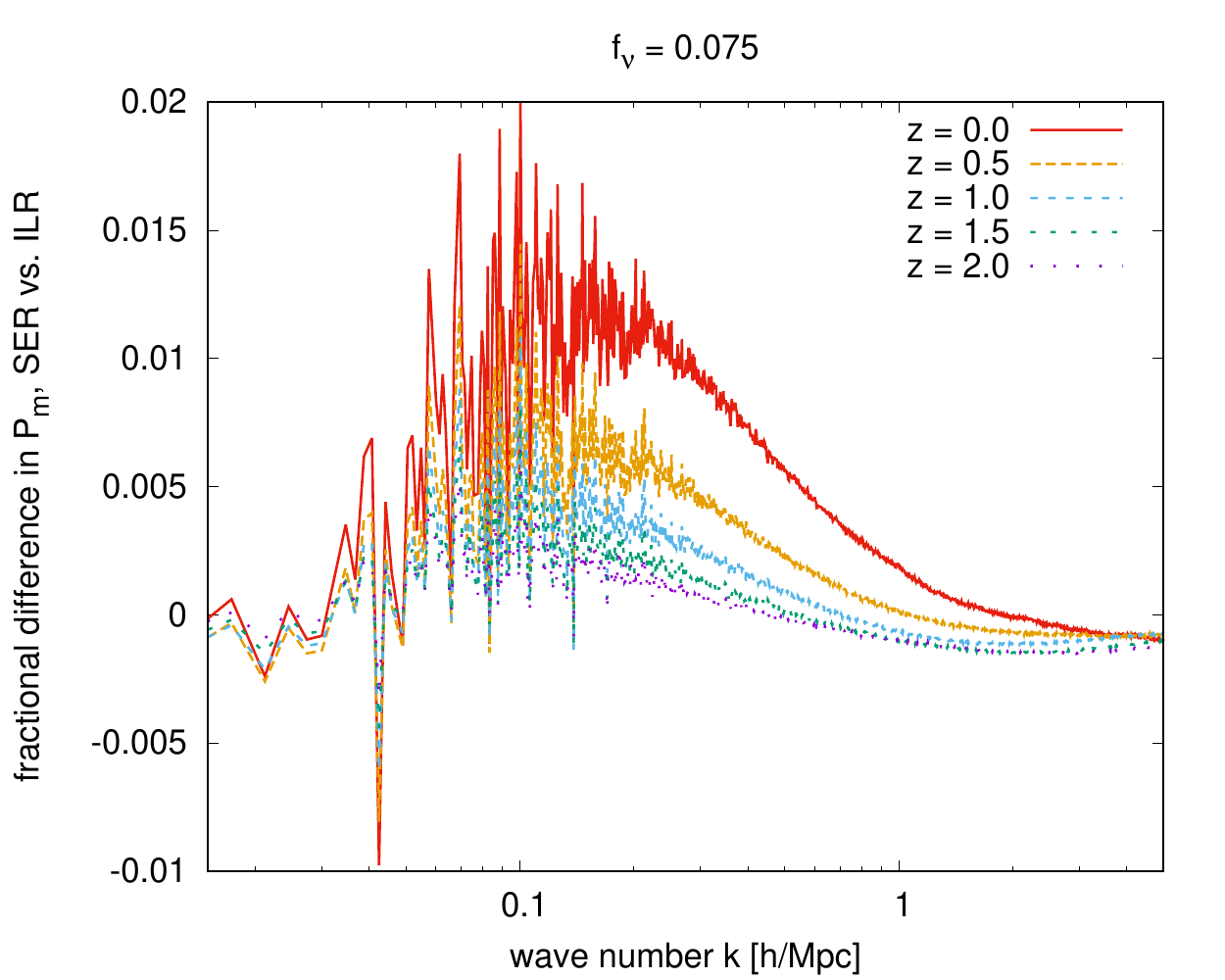}%
		\hfill
		\includegraphics[width=75mm]{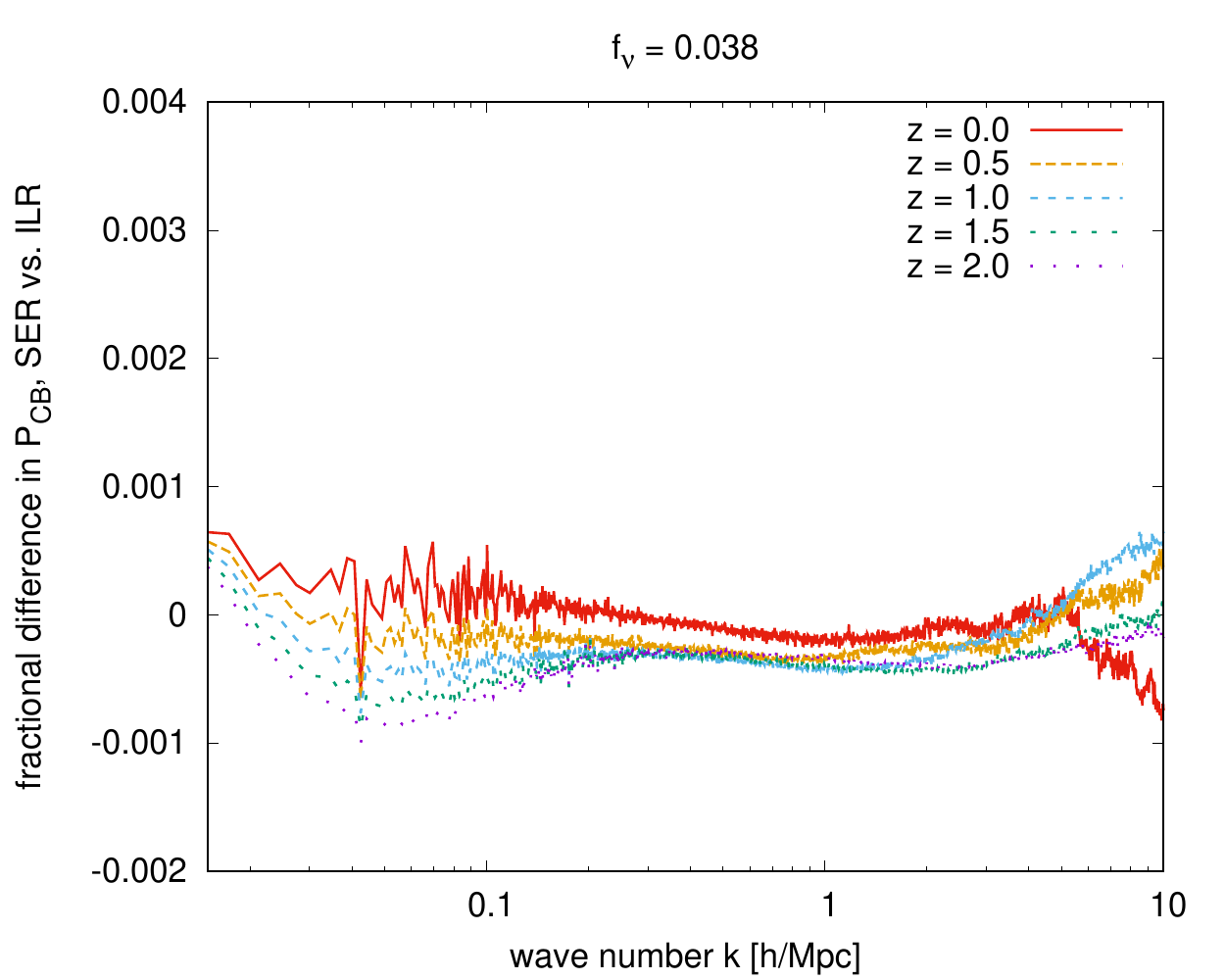}
		\includegraphics[width=75mm]{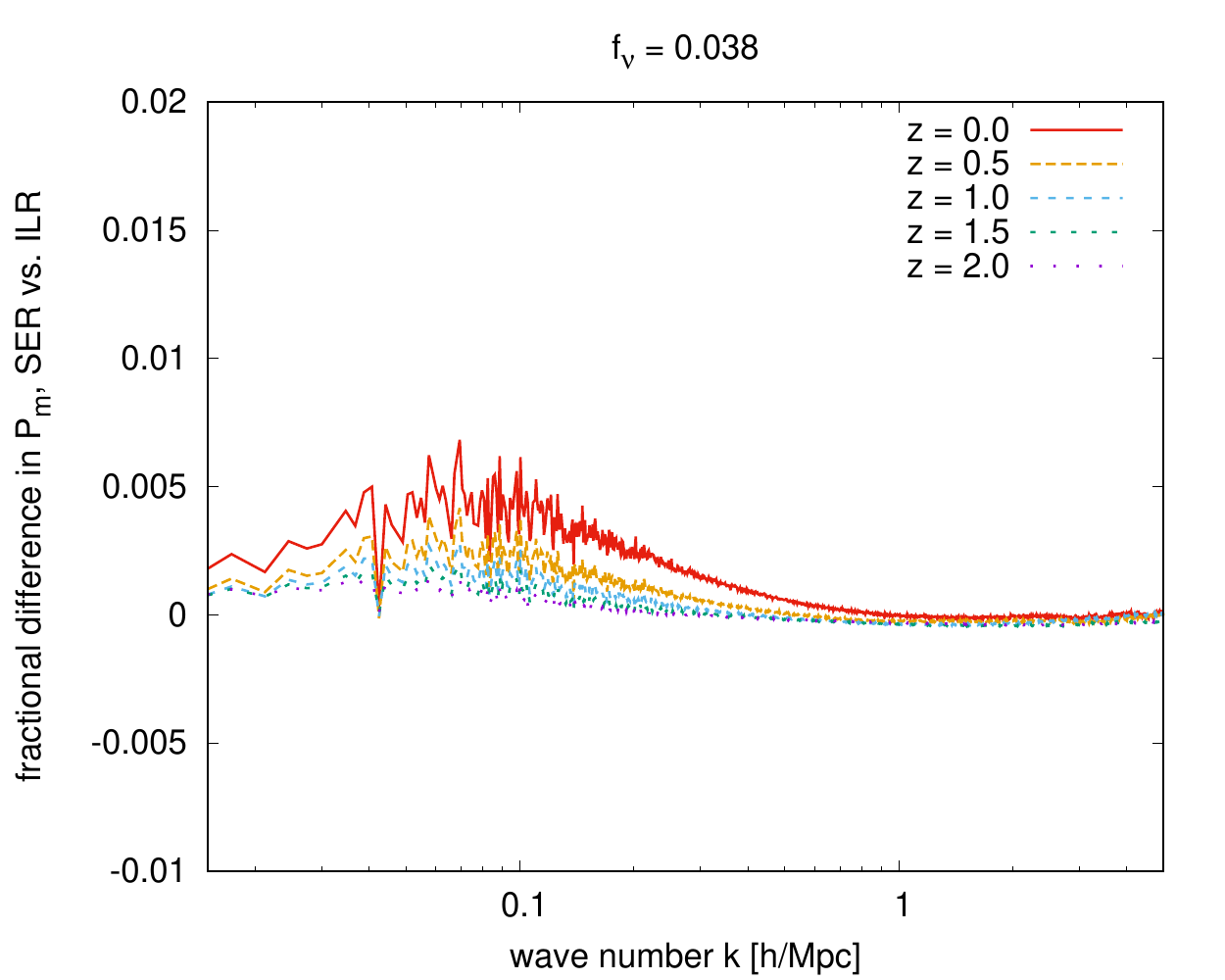}%
		\hfill
		\includegraphics[width=75mm]{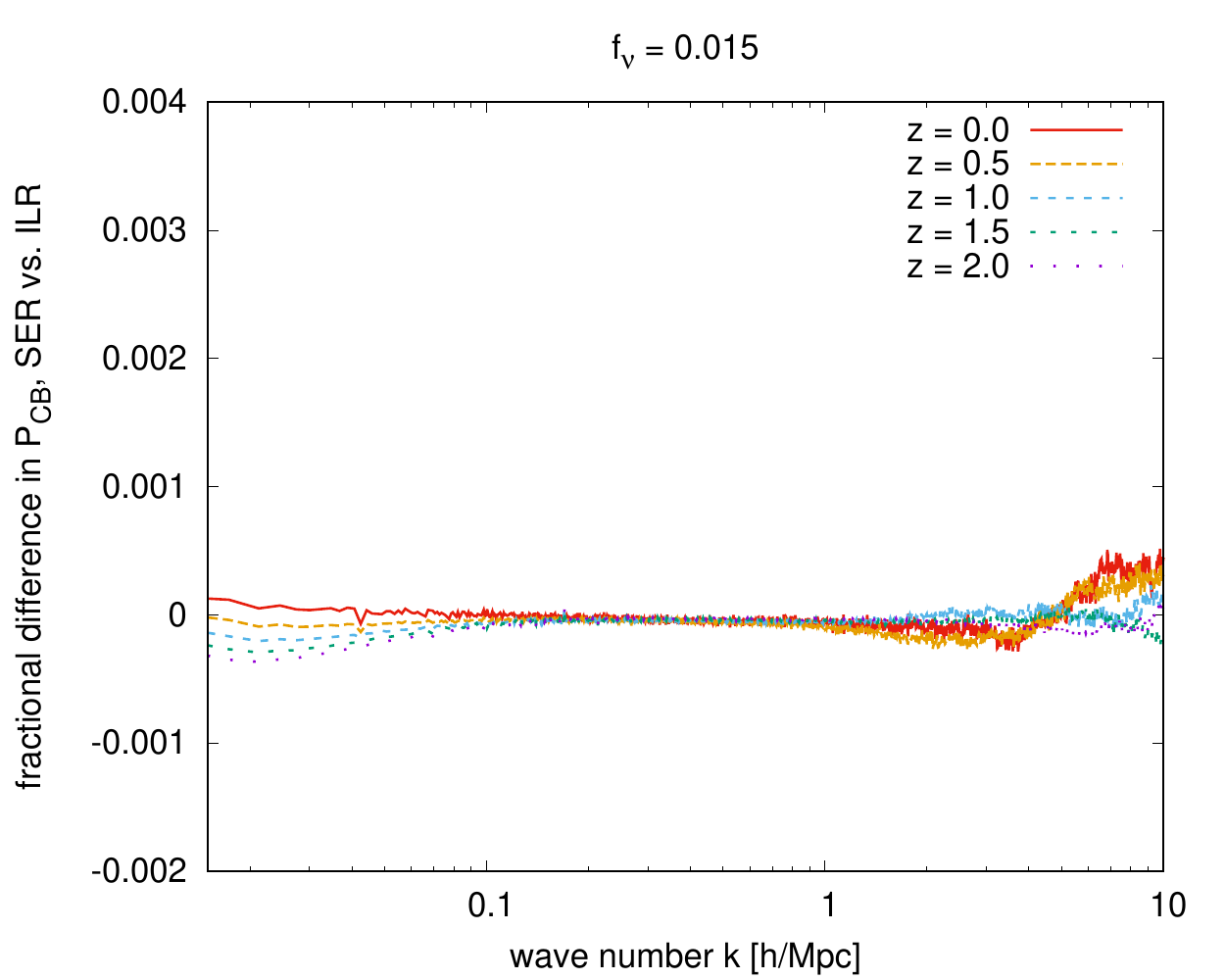}
		\includegraphics[width=75mm]{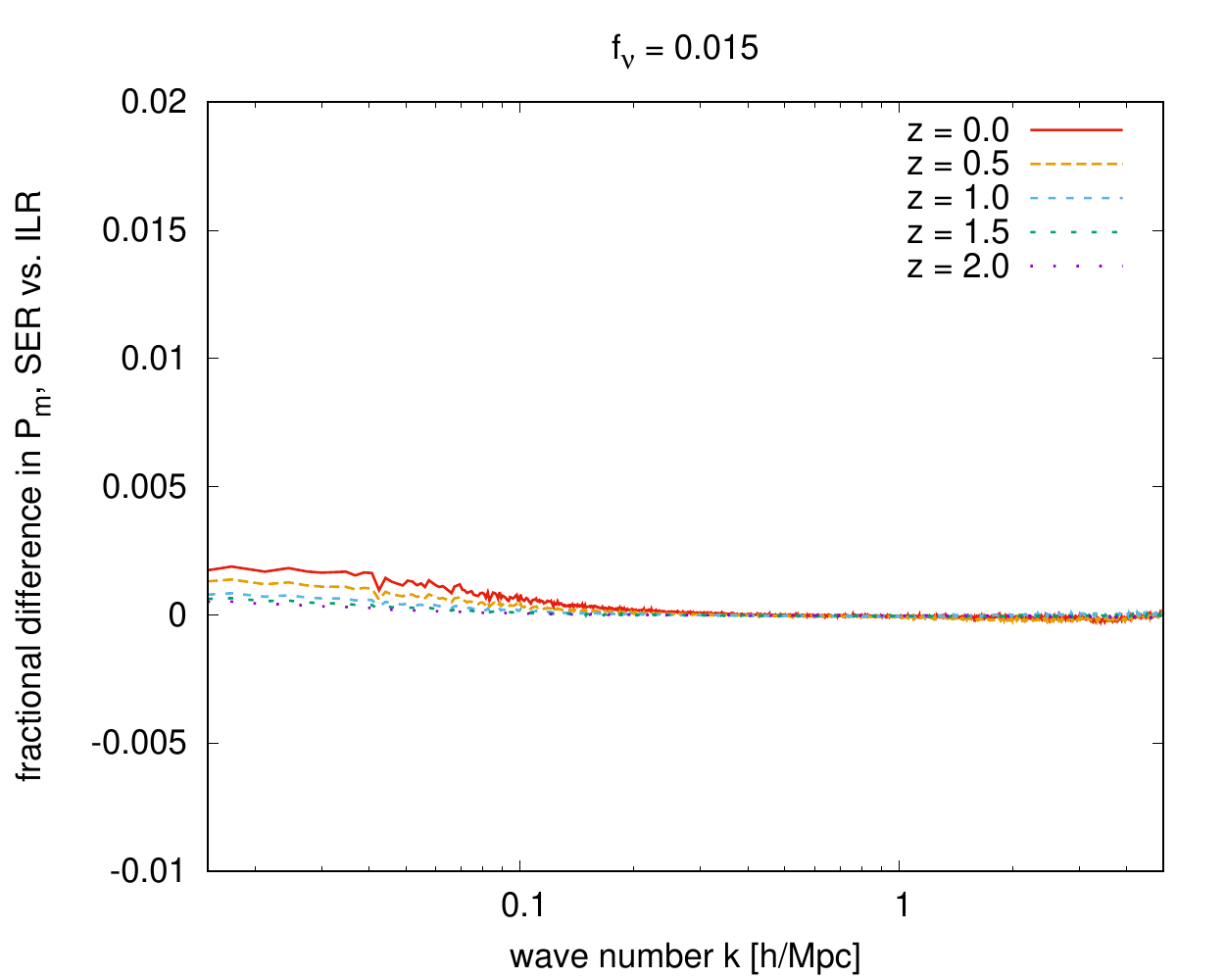}%
	\end{center}
	\caption{Fractional differences between the predictions of SuperEasy linear response (SER) and integral linear response (ILR) simulations, for cosmological models Nu1 (top row), Nu2 (middle row), and Nu3 (bottom row) at a range of redshifts.
	  {\it Left column}:  Differences in the cb power spectra.
		{\it Right column}: Differences in the total matter power spectra. The bump at $k \simeq 0.1$, which is most prominent in larger-mass models at low redshifts, can be attributed to interpolation errors in the SuperEasy response function.}
	\label{Fig:PowerComparison}
\end{figure}

Figure~\ref{Fig:PowerComparison} shows the fractional differences in the cb power spectrum and in the total matter power spectrum between the SuperEasy and integral linear response methods for the models Nu1, Nu2, and Nu3 at a range of redshifts.  Evidently, the agreement between the two methods in predicting the cb power spectrum is excellent across the whole redshift range tested (0.1\% or better), even for the most extreme Nu1 model.
  The total matter power spectrum predictions likewise agree to within sub-percent level on all scales at $z>0.5$.  Unsurprisingly, interpolation error in the SuperEasy gravitational potential causes the SuperEasy power spectrum to overshoot its integral response counterpart  around $k \sim 0.1 \, h/\text{Mpc}$  by up to 1\% at $z=0$.  As the neutrinos' contribution to nonlinear clustering diminishes with lower neutrino masses, however, the agreement between the two methods also improves.

Thus, to conclude this subsection, even in a cosmology with a neutrino mass sum as large as $\sum m_\nu \sim 1$~eV, SuperEasy linear response is able to match the original integral linear response method~\cite{AliHaimoud:2012vj} to 0.1\% agreement in the cb power spectrum and 1.2\% agreement in the total matter power spectrum. 
For a lower mass sum of $\sum m_{\nu} \sim 0.5$~eV, 
the agreement improves to 0.05\% in the cb power and 0.5\% in the total matter power. For mass sums comparable to current cosmological  bounds, $\sum m_{\nu} \sim 0.2$~eV, 0.1\% agreement can be achieved in both the cb and the total matter power spectrum across the entire accessible $k$-range.

\subsection{Comparison with grid-based linear neutrino perturbations in \pkdgrav{}} 
\label{Sec:PKDGRAV}

\begin{figure}
	\begin{center}
		\includegraphics{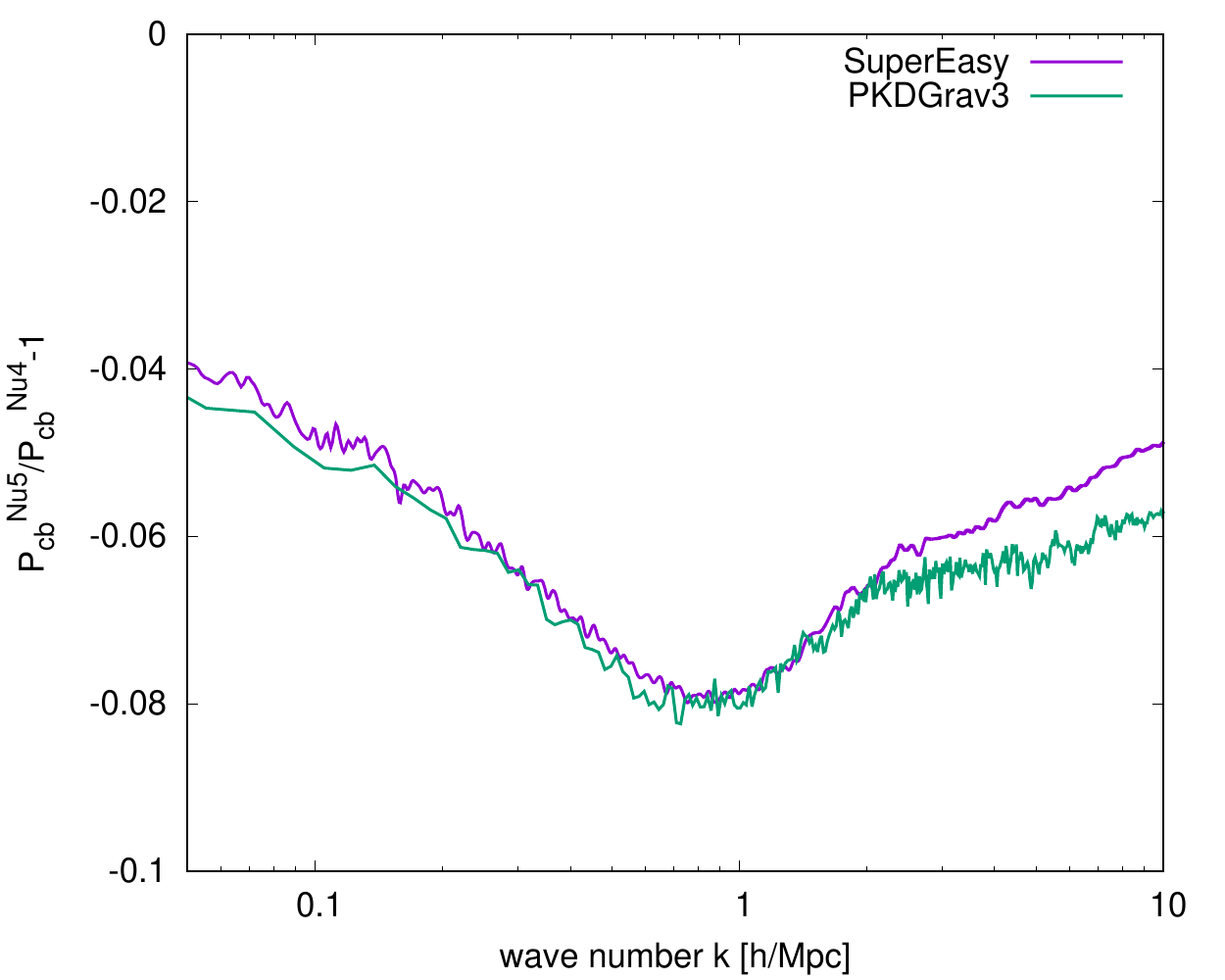}
	\end{center}
	\caption{Relative cb power spectrum between the Nu5 and Nu4 models obtained from our SuperEasy linear response simulations with~\gadgetcode{} (purple line), versus the same quantity computed independently in reference~\cite{Hannestad:2020rzl} using the grid-based linear neutrino perturbations method~\cite{Brandbyge:2008js} implemented in the \pkdgrav{} $N$-body code (green line).}
	\label{Fig:GadgetPKDComparison}
\end{figure}

Next, we compare power spectrum predictions obtained from our SuperEasy linear response \gadgetcode{} simulations
with independent calculations from reference~\cite{Hannestad:2020rzl} using \pkdgrav{}~\cite{Potter:2016ttn}. 
\pkdgrav{} is a tree code wherein the cold matter particles are evolved under gravity using a variation of the oct-tree method. Reference~\cite{Hannestad:2020rzl} simulated massive neutrinos by evolving  linear neutrino perturbations on a grid following the method of~\cite{Brandbyge:2008js}.  At every time step, the neutrino contribution is interpolated to the positions of the cold matter particles for the force calculation. 

Before we compare the two sets of results, we note first of all that a direct comparison of the {\it absolute} cb or total matter power spectrum outputted by different simulation codes for the same cosmology typically suffers from a $\sim 10$\% systematic difference~\cite{Schneider:2015yka} on small length scales,%
\footnote{The convergence ``trajectory'' between different simulation codes are not expected to be the same. Where one code may overestimate power at lower resolution, another may underestimate it. In reference~\cite{Schneider:2015yka}, this convergence-related disagreement occurs at very high $k$-values ($\sim$ 10 $h$/Mpc), as the simulations used in that work had excellent mass resolution. 
Our simulations do not have the same excellent mass resolution; the code discrepancies therefore also show up in a much lower $k$-region ($\sim$ 2 $h$/Mpc).} 
too large for the kind of comparisons we are interested in --- comparisons that concern  percent-level effects ---  to be meaningful.
Conversely, the {\it  relative} power spectrum between two cosmologies computed  under the same conditions and with the same simulation code generally end up quite insensitive to those exact conditions or the code with which the simulations have been performed, 
 because by forming ratios multiplicative systematic simulation errors tend to cancel~\cite{Hannestad:2019piu}.  We therefore choose to compare  relative power spectra.

Figure~\ref{Fig:GadgetPKDComparison} shows the $z=0$ relative cb power spectrum of the Nu5 to the Nu4 model of table~\ref{Table:M000nu1Params} obtained from our SuperEasy \gadgetcode{} simulations, alongside the same quantity formed from the \pkdgrav{} simulations of reference~\cite{Hannestad:2020rzl} for two 
equivalent massive neutrino cosmologies (labelled $\nu \Lambda$CDM5 and $\nu \Lambda$CDM4 in that work). Note that for a consistent comparison with $\nu \Lambda$CDM5 and $\nu \Lambda$CDM4, we have also simulated our models Nu4 and Nu5 using the same unequal neutrino masses.\footnote{The specific neutrino mass values in the $\nu \Lambda$CDM4 and 
$\nu \Lambda$CDM5 cosmologies  of reference~\cite{Hannestad:2020rzl} 
are, respectively,  $\{m_1, m_2, m_3\} =\{0,0.0087, 0.05 \}$~eV and $\{0, 0.0087, 0.15\}$~eV.}
See appendix~\ref{Sec:ExtHierarchy} for the details on how to implement the SuperEasy linear response method in scenarios involving multiple free-streaming lengths.
Evidently, the two sets of simulations generally agree very well with one another across the whole $k$-range tested. We find however the spoon ratio in the high-$k$  region ($k \gtrsim 2 h/\text{Mpc}$)  to be highly sensitive to the random seed used to generate a particular realisation.  Averaging over multiple realisations improves the small discrepancy  at $k \gtrsim 2 h$/Mpc seen in figure~\ref{Fig:GadgetPKDComparison}.

Thus, to conclude, our implementation of SuperEasy linear response in \gadgetcode{} is able to accurately reproduce existing independent calculations of the relative cb power spectrum between two massive neutrino cosmologies
 that not  only utilised a completely different modelling of the neutrino perturbations, but also a different $N$-body code. 
 Together with the comparison with integral linear response in section~\ref{Sec:ComparisonILR}, 
 this agreement lends credence to the SuperEasy linear response method
 as an extremely low-cost and yet reliable way to model the effects of massive neutrinos in nonlinear large-scale structure formation.


\section{Neutrino phase space distribution and the validity of linear response}
\label{subsec:neutrino_density_contrast_from_linear_response}

In the last section of this work, we test the validity of linear response as a model of neutrino clustering. A commonly-used, rule-of-thumb approach to assessing the validity of linearisation is to verify that the dimensionless power spectrum $\Delta^2(k)  = k^3 P_\nu (k) / (2\pi^2)$ of the neutrino density contrast~$\delta_{\nu}$ satisfies $\Delta^2(k) \lesssim 0.1$.  This test, while very straightforward, is unfortunately too simplistic for our purposes because of the neutrino population's large velocity dispersion, which may allow  a substantial, slowly-moving  fraction of neutrinos to cluster strongly even though most of the fast-moving ones do not.  

Indeed, what enables the linear response formalism is the approximation~\eqref{eq:linearisation}, which is a statement on the {\it perturbed neutrino phase space density} $f(\vec{k}, \vec{p}, s)$ following from the linearisation condition~\eqref{eq:linearisationcondition}. Our aim in this section, therefore, is to examine $f(\vec{k}, \vec{p}, s)$ itself in its linear response to nonlinear cold matter dynamics.  We begin by looking at how $f(\vec{k}, \vec{p}, s)$ deviates from the relativistic Fermi--Dirac distribution of the homogeneous and isotropic background, followed by an assessment of the validity of the linearisation condition~\eqref{eq:linearisationcondition}.


\subsection{Deviation from the relativistic Fermi--Dirac distribution}
\label{subsec:df_free-streaming_and_clustering_limits}

We are interested in the deviation of the perturbed neutrino phase space density from the background distribution, $\delta f(\vec{k},\vec{p},s) \equiv f(\vec{k},\vec{p},s) - \bar{f}(p)$. This can be extracted from the perturbed part of \eqref{Eq:GenSolution},
\begin{equation}
\delta f(\vec{k},\vec{p},s)
=
i m_{\nu}  k \mu \frac{\partial \bar{f}}{\partial p} \int_{s_{\rm i}}^s \mathrm{d}s' \, a^2(s') \, {\phi}(k,s') \, \text{exp} \! \left(-\frac{i k p \mu}{m_{\nu}} (s-s') \right),
\label{eq:df}
\end{equation}
where we have defined $\mu \equiv \hat k \cdot \hat p$. To keep the calculation tractable we  also assume $\phi(k,s)$ to depend only on the magnitude but not the direction of $\vec{k}$.
Define the average $\langle X \rangle_\mu \equiv (1/2) \int_{-1}^{1} {\rm d} \mu \, X(\mu)$. Applying it to equation~\eqref{eq:df} we find
\begin{equation}
  \langle \delta f \rangle_\mu (k,p,s)
 =
  m_\nu k \frac{\partial \bar f}{\partial p}
  \int_{s_{\rm i}}^s \mathrm{d}s' \, a^2(s')\, \phi(k,s') \, W\left(\frac{kp(s-s')}{m_\nu}\right),
  \label{eq:df_sprime}
\end{equation}  
with 
\begin{equation}
W(x) \equiv \frac{\sin(x)}{x^2} - \frac{\cos(x)}{x} =
- \frac{\mathrm{d}}{{\rm d}x} {\rm sinc}(x),
\label{eq:wu}
\end{equation}
where ${\rm sinc}(x) \equiv \sin(x)/x$.

As with $\delta_\nu$ in section~\ref{Sec:Superfast}, equation~\eqref{eq:df_sprime} can be solved analytically in the clustering (small~$k$) and the free-streaming (large~$k$) limits. 
 In the clustering limit, we formally set $k = 0$ in the argument of the function~$W$, and note that $W(x) \to x/3$ as $x \to 0$.  This allows us to approximate 
equation~\eqref{eq:df_sprime} in the clustering limit as
\begin{equation}
  \left< \delta f \right>_\mu^{\mathrm{C}} \simeq
  \frac{k^2}{3}  \, \frac{\partial \bar f}{\partial \ln p}
  \int_{s_{\rm i}}^s \mathrm{d}s'\,  a^2(s') \, \phi(k,s')\,  (s-s').
\end{equation}
Contrasting this expression with the clustering solution~\eqref{Eq:DeltaNuClusteringCDM} for $\delta_\nu$, we immediately find
\begin{equation}
\begin{aligned}
  \langle \delta f \rangle_\mu^{\mathrm{C}}
 \simeq
  -\frac{1}{3} \frac{\partial \bar f}{\partial \ln p}
  \delta_\nu(k,s)
   \simeq
  -\frac{1}{3} \frac{\partial \bar f}{\partial \ln p}
  \delta_{\rm m}(k,s),
  \label{eq:df_clustering}
 \end{aligned} 
\end{equation}
where the second, approximate equality follows from equation~\eqref{Eq:ClusteringDeltaNuSolution}.

The free-streaming solution, on the other hand, can be obtained by first  integrating equation~\eqref{eq:df_sprime} by parts,
\begin{equation}
\langle \delta {f} \rangle_\mu =   \frac{m_\nu^2}{p^2}   \frac{\partial \bar{f}}{\partial \ln p} \left[ \left. a^2(s') \, {\phi}(k,s') \,  {\rm sinc} \left(\frac{k p (s - s')}{m_\nu} \right)\right|_{s_i}^s  
- \int_{s_{\rm i}}^s \mathrm{d}s' \, \frac{{\rm d} (a^2 \phi)}{ {\rm d} s'} \, 
{\rm sinc} \left( \frac{k p (s - s')}{m_\nu}\right) \right].
\label{eq:byparts}
\end{equation}
 Then, demanding
analogously to equation~\eqref{eq:condition} that
\begin{equation}
\frac{1}{a^2 \phi}\frac{{\rm d} (a^2 \phi)}{ {\rm d} s} \frac{m_\nu}{k p} \ll 1,
\end{equation}
the second term in equation~\eqref{eq:byparts} can be eliminated to yield the free-streaming solution
\begin{equation}
\begin{aligned}
\langle \delta {f} \rangle_\mu^{\rm FS}& \simeq \frac{m_\nu^2}{p^2}   \frac{\partial \bar{f}}{\partial \ln p} a^2(s) \, {\phi}(k,s)
= - \frac{1}{3}   \frac{\partial \bar{f}}{\partial \ln p}  
 \frac{k_{{\rm FS},p}^2(s)}{k^2} \delta_{\rm m}(k,s),
\label{eq:df_free-streaming}
\end{aligned}
\end{equation}
where we have at the second equality used the Poisson equation~\eqref{eq:poisson}, and defined a $p$-dependent free-streaming wave number and sound speed,
\begin{eqnarray}
  k_{{\rm FS},p}(s)
 & \equiv & \sqrt{
  \frac{(3/2) \, {\mathcal H}^2(s) \, \Omega_{\rm m}(s)}{c_p^2(s)}}
  \label{eq:k_FS_v}, \\
c_p (s)& \equiv & \frac{p}{\sqrt{3} \, a m_\nu},  
\end{eqnarray}
similar to $k_{\rm FS}(s)$ and $c_\nu(s)$ of equations~\eqref{Eq:kFs} and~\eqref{Eq:FreeStreamingScale}.

\begin{figure}[t]
	\begin{center}
		\includegraphics[width=150mm]{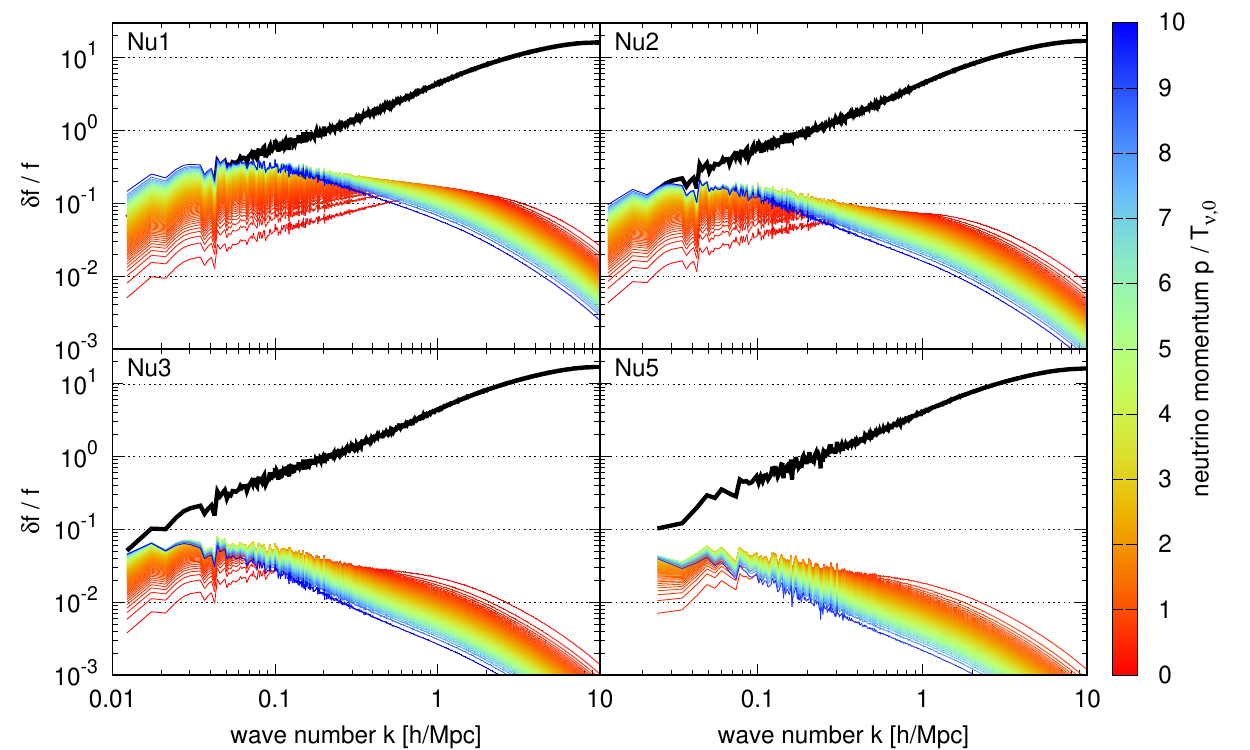}
	\end{center}
	\caption{Fractional deviation of the perturbed neutrino phase density from the background relativistic Fermi--Dirac distribution, $\langle \delta f \rangle_\mu/\bar{f}$, at $z=0$ in 100 equal-population momentum-bins, constructed from our SuperEasy linear response simulations	for the cosmological models Nu1, Nu2, Nu3, and Nu5.   Red represents the lowest-momenta, while blue denotes the highest.  For reference we also plot in black the matter density contrast $\delta_{\rm m}$, taken to be the square root of the dimensionless power spectrum $\Delta^2_{\rm m}$.
		\label{f:df}}
\end{figure}

\begin{figure}[t]
 	\begin{center}
 		\includegraphics[width=150mm]{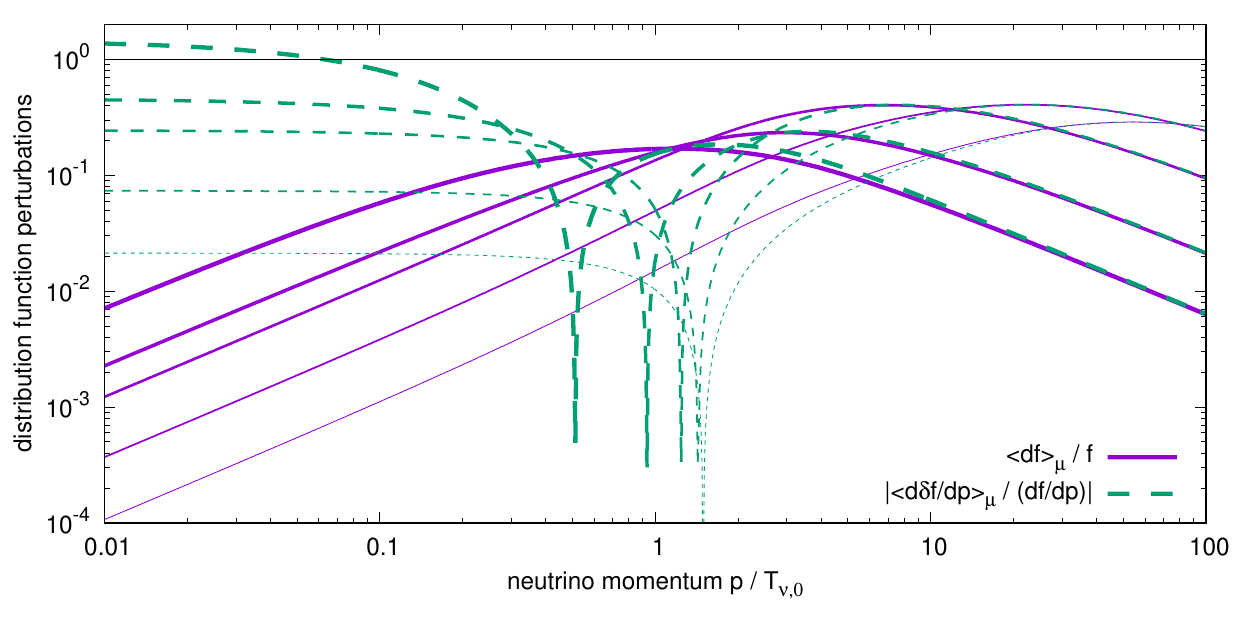}
 	\end{center}
 	\caption{Fractional deviation $\langle \delta f \rangle_\mu/\bar{f}$ (solid purple) and the quantity $I(k,p,s) = \left|(\partial \langle \delta f \rangle_\mu/\partial p) /( \partial \bar{f}/\partial p) \right|$ (dashed green) at $z=0$ for the model Nu1.
 		Each set of lines corresponds to $k = 0.01 h$/Mpc, $0.03 h$/Mpc, $0.1 h$/Mpc, $0.3 h$/Mpc, and $1 h$/Mpc in order of increasing thickness. Momenta below $0.23$, $0.51$, and $1.2$ in units of the  present-day neutrino temperature correspond to the slowest 0.1\%, 1\%, and 10\% of neutrinos, respectively, while those above $5.5$, $8.5$, and $11$ correspond to the fastest 10\%, 1\%, and 0.1\%, respectively.
 		\label{f:dfddf}}
 \end{figure}

Connecting the two limiting solutions~\eqref{eq:df_clustering} and \eqref{eq:df_free-streaming} with an interpolation function analogous to equation~\eqref{eq:interpolatem} then yields an estimate for the fractional deviation of the perturbed neutrino phase space density from the background relativistic Fermi--Dirac distribution,
\begin{equation}
 \frac{ \langle \delta f \rangle_\mu}{\bar{f}} (k,p,s)
  =
  -\frac{1}{3} \frac{\partial \ln \bar f}{\partial \ln p}
  \frac{k_{{\rm FS},p}^2(s)}{[k + k_{{\rm FS},p}(s)]^2} \, \delta_{\rm m}(k,s).
  \label{eq:df_fit}
\end{equation}
Figure~\ref{f:df} shows $\langle \delta f \rangle_\mu/\bar{f}$ in 100 equal-population momentum-bins at $z=0$, constructed from our SuperEasy linear response simulations for several massive neutrino cosmologies of table~\ref{Table:M000nu1Params}.
In line with expectations, the larger the neutrino fraction $f_\nu$ and hence more massive the neutrinos, the larger the fractional deviation of the perturbed neutrino phase space density from the Fermi--Dirac distribution.  In the case of the model Nu1, the fractional deviation can be as large as $0.4$, which may be hinting at a potential failure of linear response as a model of neutrino clustering for neutrino masses as large as $\sum m_\nu = 0.93$~eV (see also figure~\ref{f:dfddf}). The model Nu2 ($\sum m_\nu = 0.465$~eV) likewise exhibits deviations exceeding $0.1$.  Smaller-mass models however generally show no cause for concern.

Interestingly, and perhaps also counter-intuitively, it is not only at the low momenta that the neutrino phase space density deviates significantly from the background distribution. In fact, given a cosmological model, figures~\ref{f:df} and~\ref{f:dfddf} show that 
{\it all} momenta low and high can exhibit a sizeable departure from the Fermi--Dirac distribution at some wave number~$k$.  This is a feature of the Eulerian picture, and
comes about because on every length scale $R \sim 1/k$, gravitational infall causes the bulk of the neutrinos to congregate around a  momentum $p_{\rm esc} = m_\nu v_{\rm esc}$ corresponding to the escape velocity of the structures on that scale, $v_{\rm esc} = \sqrt{2 G M/R}$, where 
 $M = (4 \pi/3) \bar{\rho}_{\rm m} R^3$.  Thus, for a fixed neutrino mass, the highest momenta peak first at small wave numbers $k$, while the lowest momenta peak at large $k$ values.  In the same vein, for a fixed momentum, the smaller the neutrino mass, the smaller the wave number $k$ at which we expect a peak.  
 
 Our result is consistent with the findings of reference~\cite{Ringwald:2004np}, which likewise concluded that the largest deviation in the phase space density of relic neutrino background in the solar neighbourhood occurs at $p_{\rm esc}$ (see figure 6 of~\cite{Ringwald:2004np}).  The Eulerian picture of this work also forms 
   an interesting contrast to the semi-Lagrangian, multi-fluid linear response approach investigated in our companion paper 2~\cite{Chen:2020}, wherein neutrinos with low Lagrangian momenta do consistently exhibit a larger departure from their background densities than do the higher-momentum ones.  The two approaches are therefore complementary.


\subsection{Condition of linearisation}
\label{subsec:validity_of_f_linearisation}

Next, we turn to assessing how well the linearisation condition~\eqref{eq:linearisationcondition} is satisfied by our SuperEasy linear response simulations.
  If we are again only interested in a $\mu$-averaged result, then the condition~\eqref{eq:linearisationcondition} can be equivalently written as
\begin{equation}
I(k,p,s) \equiv \left|\frac{\partial \langle \delta f \rangle_\mu/\partial p}{\partial \bar{f}/\partial p} \right| \ll 1,
\label{eq:linearisation2}
\end{equation} 
where the numerator can be readily obtained by differentiating the expression~\eqref{eq:df_fit},
\begin{equation}
\begin{aligned}
\frac{\partial}{\partial p} \langle \delta {f} \rangle_\mu (k,p,s)  = -\frac{1}{3} \left[p \frac{\partial^2 \bar{f}}{\partial p^2} - \frac{\partial \bar{f}}{\partial p} \left(\frac{k-k_{{\rm FS},p}(s)}{k+ k_{{\rm FS},p}(s)} \right)\right] \frac{k_{{\rm FS}, p}^2(s)}{[k + k_{{\rm FS}, p}(s)]^2} \delta_{\rm m}(k,s).
\label{eq:ddfdp}
\end{aligned}
\end{equation}
Alternatively, a more laborious route that leads to the same outcome begins with $\delta f(\vec{k},\vec{p},s)$ of equation~\eqref{eq:df}. Forming from it~$\hat{p} \cdot \nabla_{\vec{p}}\, \delta f$ and then averaging over $\mu$ yields a $\langle \hat{p} \cdot \nabla_{\vec{p}}\, \delta f\rangle_\mu$ that has the same clustering and free-streaming solutions as $\partial\langle \delta f \rangle_\mu/\partial p$ of equation~\eqref{eq:ddfdp}.

\begin{figure}[t]
	\begin{center}
		\includegraphics[width=150mm]{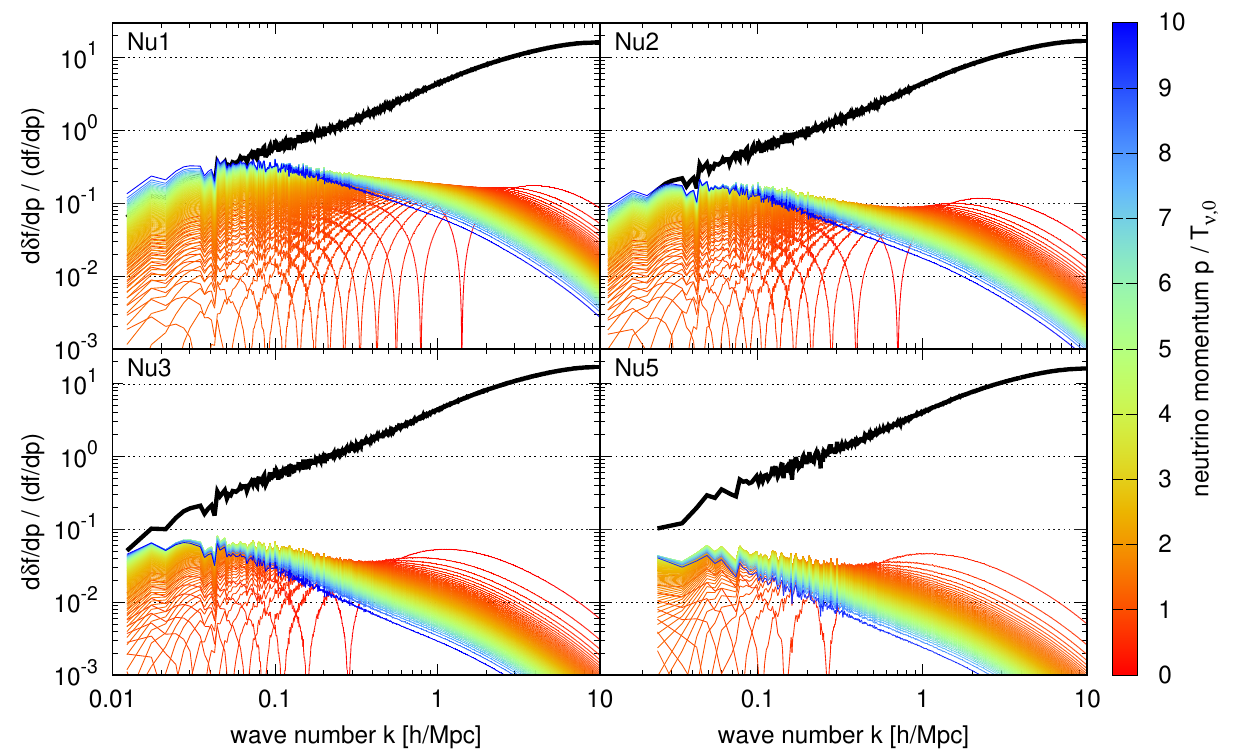}%
	\end{center}
	\caption{Same as figure~\ref{f:df}, but for the ratio $I(k,p,s)=  \left|(\partial \langle \delta f \rangle_\mu/\partial p) /( \partial \bar{f}/\partial p) \right|$.  A $I(k,p,s)$ substantially below unity indicates good justification for the method of  linear response.
		\label{f:linearity_tests}
	}
\end{figure}

Figure~\ref{f:linearity_tests} shows the ratio $I(k,p,s)$ at $z=0$ 
constructed from our SuperEasy linear response simulations
for the same four massive neutrino cosmologies as in figure~\ref{f:df}, again for 100 equal-population momentum bins.  An alternative view of the same quantity  is presented in figure~\ref{f:dfddf} for the model Nu1.  With the exception of the very lowest momenta (representing $\ll 0.1\%$ of the neutrino population) in Nu1, 
$I(k,p,s)$ remains below unity in all cases, indicating no catastrophic breakdown of the linearisation condition~\eqref{eq:linearisationcondition} or, equivalently,~\eqref{eq:linearisation2} in the massive neutrino cosmologies considered in this work.
However, as expected, the larger the neutrino mass of the cosmology, the more difficult it is to satisfy the linearisation condition.  The model Nu1 ($\sum m_\nu = 0.93$~eV), for example, has $I(k,p,s)$ reaching $0.2$ to $0.4$ in all 100 momentum-bins in figure~\ref{f:linearity_tests},  suggesting that there are likely nonlinear effects in the neutrino sector that we have failed to account for by linearisation.  The highest momentum-bins in Nu2 ($\sum m_\nu = 0.465$~eV) likewise climb to $I \sim 0.2$, although momenta in the bulk tend not to exceed $I \sim 0.1$, so that on the whole the linearisation condition~\eqref{eq:linearisationcondition} or~\eqref{eq:linearisation2} can be deemed reasonably well satisfied by the model.  The lower-mass Nu3 and Nu5 always have $I < 0.1$ in all 100 bins.

As with our discussion of $\langle \delta f \rangle_\mu/\bar{f}$ in  section~\ref{subsec:df_free-streaming_and_clustering_limits}, a unique feature of the Eulerian picture is that, for a given cosmology, all momenta tend to perform more or less the same in terms of how well they satisfy the linearisation condition~\eqref{eq:linearisationcondition} or~\eqref{eq:linearisation2}, except that the ratio $I(k,p,s)$ tends to peak at different values of $k$ for different values of $p$.
Again, this is in complete contrast to the semi-Lagrangian multi-fluid approach of our companion paper 2~\cite{Chen:2020}, where it is always the low Lagrangian-momentum streams that become nonlinear first, while the high-momentum modes stay safely linear.
 
This observation implies that incorporating nonlinear corrections in the SuperEasy linear response method of this work and in the multi-fluid linear response of paper 2  will require completely different strategies.  As discussed in paper 2, a natural nonlinear extension to the multi-fluid approach --- with its clear demarcation of nonlinear and linear Lagrangian-momentum streams ---  is the hybrid scheme proposed in reference~\cite{Brandbyge:2009ce}.  Extending SuperEasy linear response to a nonlinear response theory, on the other hand,  may require that we target the momenta around $p_{\rm esc}$ at each wave number~$k$.  We shall leave this investigation for a future publication, and close this section by emphasising that the linearisation condition~\eqref{eq:linearisation2} is generally well satisfied by massive neutrino cosmologies with mass sums not exceeding $\sum m_\nu \sim 0.5$~eV.  This range encompasses all neutrino mass sums lying within even the most conservative of the present generation of $2 \sigma$ cosmological neutrino mass bounds.


\section{Conclusions}
\label{Sec:Conclusions}

In this work, we have presented a new linear response method for incorporating massive neutrinos into a cosmological  $N$-body simulation. The new method, dubbed ``SuperEasy linear response'', builds upon analytical solutions of the linearised collisionless Boltzmann equation for the neutrino density contrast in the free-streaming and clustering limits. The two solutions are connected by an interpolation function formed of a rational function of the wave number $k$, with coefficients given by simple algebraic expressions of the total matter density $\Omega_{\rm m,0}$, the scale factor~$a$, and the neutrino mass~$m_\nu$.  The end result is an elegant one-line modification to the gravitational potential
capable of reproducing the effects of massive neutrinos from the cold matter density contrast alone, that can be immediately incorporated into the Particle--Mesh component of an $N$-body code at no additional implementation cost.

Implementing this one-line modification in \gadgetcode{}, we demonstrate that for massive neutrino cosmologies with neutrino mass sums not exceeding  $\sum m_\nu \simeq 0.9$~eV, the SuperEasy method is able to reproduce the total matter power spectrum predictions in the wave number range $0.01\, h/{\rm Mpc}\lesssim k \lesssim 5\, h/{\rm Mpc}$ to sub-percent level agreement with the state-of-the-art integral linear response method of reference~\cite{AliHaimoud:2012vj}.  For the combined cold dark matter+baryon power spectrum, our calculations are within $0.1\%$ of the state-of-the-art in the same mass and wave number range.
We further compare  the cold matter power spectrum ratio of two different massive neutrino cosmologies (the ``relative power spectrum'') computed with the SuperEasy method using \gadgetcode{}, with the equivalent result of reference~\cite{Hannestad:2020rzl} which used the grid-based linear neutrino perturbation method~\cite{Brandbyge:2008js} implemented in~\pkdgrav{}~\cite{Potter:2016ttn}.  Again, sub-percent level agreement is found.
 
We then turn to assessing the validity of linear response methods, by examining (i)~the fractional deviation of the perturbed neutrino phase space distribution from the background relativistic Fermi--Dirac distribution, and (ii)~the momentum-gradient of the deviation, which is a measure of how well the linearisation condition~\eqref{eq:linearisationcondition} is satisfied.  We find that the lower-mass models ($\sum m_\nu \lesssim 0.5$~eV) can be deemed safely linear, suggesting that the neutrino density contrast returned by SuperEasy linear response in these cases are trustworthy.  The largest-mass model considered ($\sum m_\nu \simeq 1$~eV), on the other hand, likely suffers from nonlinear effects in the neutrino sector not captured by linearisation.  We leave the investigation of nonlinear corrections  within the SuperEasy approach for a future publication.

Of course, notwithstanding its simplicity and ability to predict some cosmological observables with good accuracy, the SuperEasy method is not without limitations.  In particular, in using an interpolation function to connect the neutrino overdensities~$\delta_{\nu}$ at small and large wave numbers, we sacrifice precision in the prediction of $\delta_{\nu}$ even on nominally linear scales ($k \lesssim 0.1 \, h$/Mpc at $z=0$)
relative to the \classcode{} output.  Furthermore, as with all $N$-body methods that employ some form of linearisation for the neutrino component,
 higher-order correlation statistics such as the matter bispectrum or any other observable that probes the relative phase between the cold matter and the neutrino perturbations may not be accurately captured~\cite{Fuhrer:2014zka}.  These limitations need to be heeded when applying the SuperEasy method.

To conclude, the SuperEasy linear response method has an extremely slim profile in terms of computational cost, and is, importantly, very simple  to implement into a wide range of $N$-body codes.   The memory cost of a single SuperEasy simulation is identical to a $\Lambda$CDM simulation, with no additional storage or post-processing required. Notwithstanding its extreme simplicity, the SuperEasy method is able to match state-of-the-art massive neutino linear response methods to sub-0.1\% agreement on the cold matter power spectrum. It is an especially appealing method where limited computational resources need to be diverted to the implementation of 
other physical calculations, such as the inclusion of baryonic physics via hydrodynamics.


\acknowledgments

JZC acknowledges support from an Australian Government Research Training Program Scholarship.
AU and  Y$^3$W are  supported by the Australian Research Council's Discovery Project (project DP170102382) and Future Fellowship (project FT180100031) funding schemes.   
This research includes computations using the computational cluster Katana supported by Research Technology Services at UNSW Sydney. We thank J.M. Dakin for useful discussions.

\appendix

\section{Generalisation to multiple non-degenerate neutrino species}
\label{Sec:ExtHierarchy}

It is straightforward to generalise the SuperEasy linear response interpolation function~\eqref{Eq:NeutrinoDensityContrastResult} for the neutrino density contrast and hence~\eqref{Eq:MatterDensityContrastResult} for the total matter density to accommodate multiple species of non-degenerate neutrinos.  

Consider $N=3$ neutrino species, where each individual species~$i$ is characterised by a free-streaming scale $k_{{\rm FS},i}(s)$ and mass 
fraction~$f_{\nu_i} \equiv \bar{\rho}_{\nu_i}/\bar{\rho}_{\rm m}$.  Because the linear response solution~\eqref{Eq:MasterNeutrinoEquation} concerns only how one species responds to a formally {\it external and unspecified} potential, the limiting solutions~\eqref{Eq:FreeStreamDeltaNuSolution} and~\eqref{Eq:ClusteringDeltaNuSolution} remain valid, and we can immediately generalise the interpolation function~\eqref{eq:interpolatem} to

\begin{equation}
\begin{aligned}
{\delta}_{\nu_i}(\vec{k}, s) = & \, \frac{k_{\text{FS},i}^2(s)}{\left[k+k_{\text{FS},i}(s)\right]^2} \, {\delta}_{\rm m}(\vec{k},s) \\
= & \, \frac{k_{\text{FS},i}^2(s)}{\left[k+k_{\text{FS},i}(s)\right]^2} \left[f_{\rm cb} \, {\delta}_{\text{cb}} (\vec{k},s) + \sum_{j=1}^3 f_{\nu_j} \, {\delta}_{\nu_j}(\vec{k},s)  \right] \, ,
\end{aligned} 
\end{equation} 
where the total matter density contrast ${\delta}_{\rm m}$ now contains contributions from three neutrino species, and $f_{\rm cb} = 1 - \sum_{j=1}^3 f_{\nu_j}$.    Collecting all terms for $i$th species, we arrive at 
\begin{equation}
\begin{aligned}
{\delta}_{\nu_i}(\vec{k}, s) = & \, K_i(k,s) \left[ f_{\rm cb}  {\delta}_{\text{cb}}(\vec{k},s) + \sum_{j \neq i} f_{\nu_j} {\delta}_{\nu_j} (\vec{k}, s)\right] \, ,
\end{aligned}
\label{Eq:IndividualHierarchyDeltaNu}
\end{equation}
with 
\begin{equation}
K_i(k,s) \equiv  \frac{k_{\text{FS}, i}^2(s) }{[k+k_{{\rm FS},i}(s)]^2 - k_{\text{FS}, i}^2(s)  \, f_{\nu_i}}.
\end{equation}
Save for an additional term proportional to the other neutrino species $j \neq i$,  the interpolation function~\eqref{Eq:IndividualHierarchyDeltaNu} has essentially the same form as equation~\eqref{Eq:NeutrinoDensityContrastResult}.

Defining a neutrino density contrast vector $\vec{\delta}_{\nu} (\vec{k}, s) \equiv ( {\delta}_{\nu1}, {\delta}_{\nu_2}, {\delta}_{\nu_3})^{T}$ and 
rewriting equation~\eqref{Eq:IndividualHierarchyDeltaNu} in matrix form, we find
\begin{equation}
\vec{\delta}_{\nu}(\vec{k}, s) = \left[\mathbb{1} - \mathrm{M}(k,s)\right]^{-1} \vec{\xi}(k,s) \, \, f_{\rm cb}\, {\delta}_{\text{cb}}(\vec{k},s) \, ,
\label{Eq:GeneralDeltaNuHierarchy}
\end{equation}
where  $\vec{\xi} \equiv ( K_1, K_2, K_3)^T$, $\mathrm{M}(k,s)$ is a square matrix defined as
\begin{equation}
\mathrm{M} (k,s)\equiv 
\begin{pmatrix}
0 &K_1 f_{\nu 2}& K_1 f_{\nu_3} \\
K_2 f_{\nu_1} & 0 & K_2 f_{\nu_3}\\
K_3 f_{\nu_1} & K_3 f_{\nu_2}& 0 
\end{pmatrix} \, ,
\end{equation}
and $[\cdots]^{-1}$ denotes an inverse operation.  Having thus expressed ${\delta}_{\nu_i}(k,s)$ solely in terms of the CDM--baryon density contrast~${\delta}_{\text{cb}}(\vec{k},s)$, they can now be combined to form the total matter density~${\delta}_{\text{m}}(\vec{k},s)$ and hence the gravitational potential ${\phi}$ via the Poisson equation~\eqref{eq:poisson}.  Implementation into $N$-body simulations can then proceed  as described in section~\ref{Sec:Nbody}.

\bibliographystyle{utcaps}
\bibliography{SuperEasyBib}

\end{document}